\documentclass[letter,12pt]{article} 

\usepackage[latin1]{inputenc} 
\usepackage{amssymb}
\usepackage{cite}
\usepackage{amsfonts}
\usepackage{amsthm}
\usepackage{multicol}
\usepackage{multirow}
\usepackage{amsmath}
\usepackage{graphicx}
\usepackage{float}
\usepackage{hyperref}

\begin{document}

\begin{center}
\textbf{Influence of isoscalar and isovector pairing on Gamow-Teller transitions for nuclei in the $2p1f$ shell: A schematic shell model study.} 

\vspace{0.25cm}

\textbf{A. Carranza M. $^{(1)}$, S. Pittel $^{(2)}$, Jorge G. Hirsch $^{(1)}$.} 

(1) Instituto de Ciencias Nucleares, Universidad Nacional Aut\'onoma de M\'exico, 04510, M\'exico CDMX,
M\'exico.

(2) Bartol Research Institute and Department of Physics and Astronomy,University of Delaware,
Newark, Delaware 19716, USA.
\end{center}

\section*{Abstract}

We perform a systematic study of Gamow-Teller (GT) transitions in the $2p1f$ shell, using the nuclear shell model with two schematic Hamiltonians. The use of the shell model provides  flexibility to analyze the role of different proton-neutron pairing modes in the presence of nuclear deformation. The schematic Hamiltonians that are used contain a quadrupole-quadrupole interaction as well as isoscalar ($T=0$) and isovector ($T=1$)  pairing interactions, but differ in the single particle energies. The objective of the work is to observe the behavior of GT transitions in different isoscalar and isovector pairing scenarios, together with the corresponding energy spectra and rotational properties of the parent and daughter nuclei ( $^{42}Ca \rightarrow ^{42}Sc$, $^{44}Ca \rightarrow ^{44}Sc$, $^{46}Ti \rightarrow ^{46}V$, $^{48}Ti \rightarrow ^{48}V$). We also treat the rotational properties of $^{44}Ti$ and $^{48}Cr$. All results are compared with experimental data. The results obtained from our models depend on the different scenarios that arise, whether for $ N = Z $ or $ N \neq Z$ nuclei. In the latter case, the presence of an attractive isoscalar pairing interaction imposes a $1^{+}$ ground state in odd-odd nuclei, contrary to observations for some of the nuclei considered, and it is necessary to suppress that pairing mode when considering such nuclei. The effect of varying the strength parameters for the two pairing modes is found to exhibit different but systematic effects on energy spectra and on GT transition properties.

\section{Introduction}

Gamow-Teller (GT) transitions are a very important tool in the exploration of nuclei \cite{prc971,prc972,prc975,prc976}. They have important applications in the $\beta$ decay and electron capture processes that arise in stellar evolution \cite{prc977,prc979}, in  the double capture of electrons in hot stars and in neutrino nucleosynthesis \cite{prc9713,prc9714,prc9715}, in addition to their importance for testing nuclear models.  There are two types of GT transitions;  $ GT_{+}$ in which a proton is changed into a neutron, and $ GT_{-}$ in which a neutron is changed into a proton. The transition strength B(GT) can be obtained from $\beta$ decay studies, but with excitation energies limited by the $Q$ values of the decays. On the other hand, with charge-exchange reactions (CE), such as $(p,n)$, $(n,p)$, $ (d,^{2}He) $, or $ (^{3}He,t) $, it is possible to access GT transitions for large values in energy without the $Q$ value restriction. Experimental measurements for such reactions at angles close to $ 0^{\circ} $ and with an incident energy above $100~MeV$/nucleon provide valuable information about GT transitions.

In this work, we focus on GT transition intensities in the $2f1p$ shell and particularly for nuclei in the $A=40-48$ region. For these nuclei, the transitions $GT_{+}$ and the reactions $(n,p)$ and $(d,^{2}He)$ are the primary experimental tools used to obtain the distribution of intensities, but the reaction $ (^{3}He,t) $ has served as an alternative experimental tool \cite{Molina}. The GT transition intensities obtained from these experimental probes will serve as a testing ground for the theoretical studies carried out here.

Our study of GT properties in this region will be done in the framework of the nuclear shell model and with a series of schematic Hamiltonians \cite{Stu} that include a quadrupole-quadrupole interaction as well as both isoscalar and isovector pairing interactions.  In this way, we will be able to assess the role of the different pairing modes on GT transition properties in a model in which deformation is incorporated naturally while symmetries are fully respected.

In general, proton-neutron (pn) pairing is expected to play an important role in nuclei with $ N \approx Z $ \cite{Stu1}. Proton-neutron pairing arises in two channels, the isoscalar (T=0) channel and the isovector (T=1) channel, both of which are included in our schematic analysis. Of particular interest is the isoscalar pairing channel as it is expected to be most especially important for nuclei with $N=Z$ and $ N \approx Z $ \cite{Stu5,Stu6}.

A summary of the paper is as follows. In Sec. II, we briefly describe the two models we consider and then in Sec. III describe selected results of our analysis, first for energy spectra and rotational properties and then for GT transitions.  Finally, in Sec. IV we summarize some of the key conclusions of the work.

\section{Models}

In the systematics to be studied, we focus on nuclei in which valence neutrons and protons outside a doubly-magic $^{40}Ca $ core are restricted to the orbitals of the $2p1f$ major shell. The first Hamiltonian we consider (denoted Model 1) is

\begin{equation} \label{model1}
\begin{split}
\widehat{H}_{1}=\chi \left(  \widehat{Q}\cdot \widehat{Q} + a \widehat{P}^{\dagger} \cdot \widehat{P} + b \widehat{S}^{\dagger} \cdot \widehat{S} +\alpha \sum_{i} \widehat{\overrightarrow{l}_{i}}\cdot \widehat{\overrightarrow{s}_{i}} \right).
\end{split}
\end{equation}
Here $\widehat{Q} = \widehat{Q}_{n} + \widehat{Q}_{p}$ is the quadrupole mass operator and $\widehat{Q}\cdot \widehat{Q}$ is the quadrupole-quadrupole operator, which includes both a one-body and a two-body term (often denoted $:\widehat{Q}\cdot \widehat{Q}:$). Also $\widehat{P}^{\dagger}$ is the operator that creates a correlated pair with $L=0$, $S=1$, $J=1$, $T=0$, whereas  $\widehat{S}^{\dagger}$ is the operator that creates a correlated pair with $L=0$, $S=0$, $J=0$, $T=1$. Finally, the last term is the contribution to the Hamiltonian in this model resulting from the single-particle part of the spin-orbit interaction.

We also consider a second model (denoted Model 2)  in which the single- particle energies $\varepsilon_{i}$ are taken from the realistic  interaction  $kb3$, viz:

\begin{equation} \label{model2}
\begin{split}
\widehat{H}_{2}=\sum_{i} \varepsilon_{i} \widehat{n}_i + \chi \left(  :\widehat{Q}\cdot \widehat{Q}: + a \widehat{P}^{\dagger} \cdot \widehat{P} + b \widehat{S}^{\dagger} \cdot \widehat{S} \right),
\end{split}
\end{equation}
Here we use the same two-body part of the Hamiltonian as for Model 1, but now use for the single-particle energies the values (see \cite{spe}) $\varepsilon_{7/2}=0.0MeV$, $\varepsilon_{3/2}=2.0MeV$, $\varepsilon_{1/2}=4.0MeV$ and $\varepsilon_{5/2}=6.5MeV$.

Within the context of these two model Hamiltonians, we explore the role of the two pairing modes on the various observables of interest, namely
energy spectra and GT transitions. We will also study how they impact the rotational properties of the nuclei under discussion.

All calculations reported here have been carried out using the ANTOINE shell-model program \cite{Ant1,Ant2,Ant3}.

The calculations are carried out systematically as a function of the  parameters that enter the two model Hamiltonians. We focus on the effect of varying the strength parameters for the isovector and isoscalar pairing terms, leaving the strength of the quadrupole-quadrupole interaction and the single-particle energies unchanged. We carry out the calculations for both the parent and daughter nuclei of relevance to the GT decays as well as a few other nearby nuclei of interest.

We will first consider the effects on the energy spectra of these nuclei and on the impact of the different modes of pairing on their rotational properties. The analysis is carried out for even-mass nuclei near the beginning of the $2f1p$ shell,  with $A=42-48$, and for both of the models of single-particle energies introduced above.  Finally, we study the effects of  parameter variations for these two models on GT transition probabilities.  Selected results of the various calculations are presented in the following section.

\section{Results}

A theme of the work is to first find {\it optimal} Hamiltonians for the two models, and then to vary the relevant pairing strength parameters away from these optimal values to see how these changes impact the description of the properties of interest. The {\it optimal} Hamiltonians will be chosen from three criteria:
\begin{enumerate}
\item optimal reproduction of the properties of the low-energy spectra of the nuclei of interest (and especially the $1^+$ states of odd-odd nuclei),
\item an optimal description of the energy spread of the spectrum up to relatively high energies, and
\item an optimal description of GT properties and in particular the associated fragmentation.
\end{enumerate}
Here we simply present the optimal parameters, saving for the following subsections a demonstration of how they emerge.

In the case of the Model 1 Hamiltonian (\ref{model1}), our analysis suggests that for $N=Z$ nuclei the optimal set of parameters are:  $ \chi=-0.065\ MeV$, $a=b=6 $, and $ \alpha=20$. In the case of the Model 2 Hamiltonian (\ref{model2}), we find that the same optimal parameters apply acceptably for $\chi$, $a$ and $b$, again when focusing on $N=Z$ nuclei.

In the case of $ N \neq Z $ our analysis suggests, as will be shown subsequently,  that the presence of isoscalar pairing causes an inappropriate behavior for the energies of the lowest $1^+$ states. Thus we will {\it turn off} the isoscalar pairing, (\i.e. set $a=0$), when dealing with $N \neq Z$ nuclei, keeping all other parameters precisely the same as for $N=Z$ nuclei. As we will see this will repair the inappropriate  $1^+$ behavior.

In what follows we first discuss energy spectra and then subsequently GT transitions. It should be borne in mind, however, that both sets of properties contribute, as discussed above, to our choice of optimal parameter sets.

\subsection{Energy Spectra and rotational properties}

Below are the results corresponding to the calculation of the energy spectra and associated rotational properties.

\subsubsection{ A=42}

First we analyze the energy spectra of the daughter and parent nuclei with $A=42$, to see what happens in different pairing scenarios. In Figs. 1 and 2, results for the energy spectra for the daughter nucleus $^{42}Sc$ are shown, using Models 1 and 2, respectively.  These calculations assume an optimal quadrupole strength of $\chi=-0.065 MeV$, which as we will see provides the desired optimization for all the features that we noted above were of central interest. It should be noted that this is slightly larger than the value used in earlier works \cite{Stu} in which GT transitions were not considered.

\begin{figure}[H]
\centering
\includegraphics[width=1.0\textwidth]{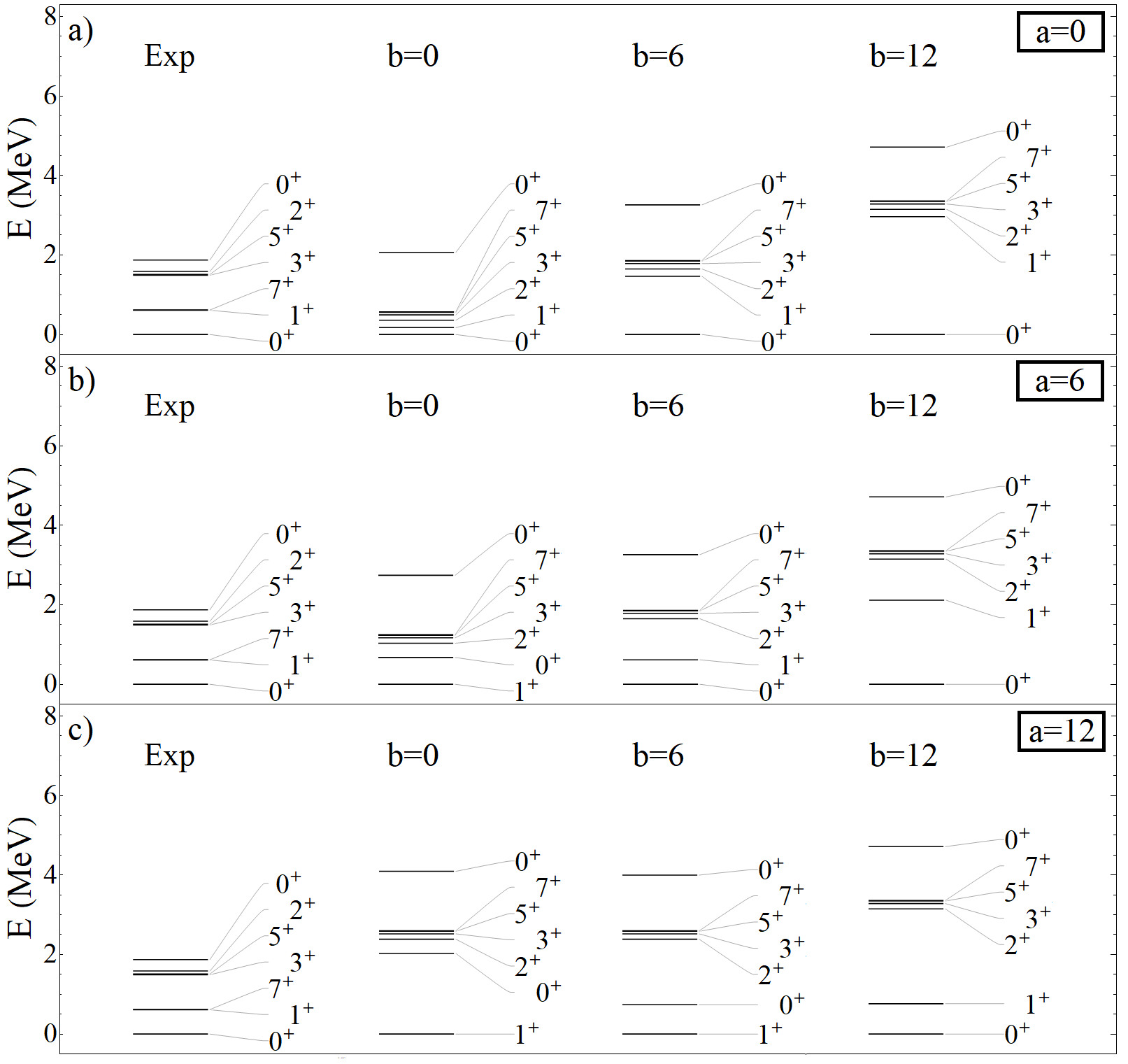}
\caption{Energy spectra of $^{42}Sc$ from Model 1 as a function of the isovector $b$ and isoscalar $a$  pairing strengths. Results are shown for $a,~b~=0,~6$, and $12$. }
\end{figure}

The results for Model 1 make use of the same spin-orbit strength, $\alpha=20$, as was used previously in those earlier works \cite{Stu}. We show results for three values of the isoscalar and isovector pairing strengths, $a = 0, 6, 12$ and $b = 0, 6, 12$, to see the effects of varying them. We will follow this strategy of spanning the optimal values in all of the results to follow.

Two points should be noted: (1) The best overall reproduction of the experimental spectrum occurs for equal values of $a$ and $b$, and (2) the optimal choice is $a=b=6$, as noted in the previous section.  With these values we produce the lowest $1^{+}$ state almost exactly at the correct energy and a reasonable density of low-lying states. The results though optimal for this Hamiltonian parametrization still show clear limitations of the model. In particular, it is not possible to produce a very low-lying $7^{+}$ state, as is present in the data.

Next we consider the results illustrated in Fig. 2 for the Model 2 Hamiltonian. Here we see again that it is necessary to have both pairing modes present to achieve a reasonable fit to both the $1^+$ energy and the density of states in the low-energy spectrum.  While it may be possible to slightly improve the results by permitting slightly different $a$ and $b$ values for this nucleus, we will maintain the same common value of $a=b=6$ as for Model 1.  However, it is  important to note again that even with these more realistic single-particle energies, we cannot reproduce the very low energy of the $7^+$ state. It is necessary to include other components of the two-body interaction to reproduce this. We will, however, maintain the current structure of the Hamiltonians in the analysis to follow, as it is the goal of this work to isolate those effects related solely to isoscalar and isovector pairing.

\begin{figure}[hbtp]
\centering
\includegraphics[width=1.0\textwidth]{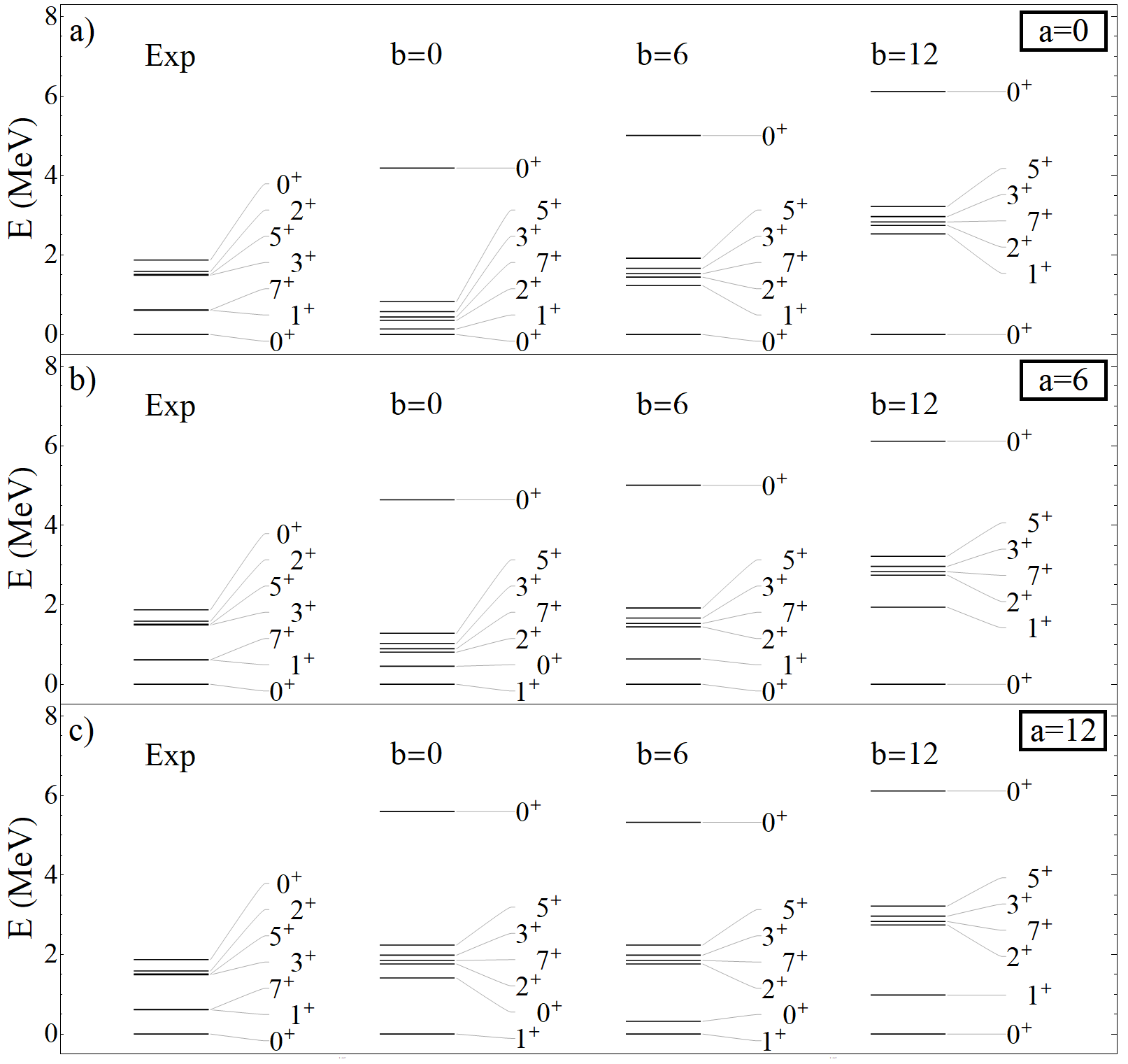}
\caption{Energy spectra of $^{42}Sc$ from Model 2 as a function of the isovector $b$ and isoscalar $a$  pairing strengths. Results are shown for $a,~b~=0,~6$, and $12$. }
\end{figure}

Another point worthy of note in Figs. 1 and 2 concerns the results in the presence of pure isovector pairing. For both models, when the isovector pairing strength $b$ is weak the ground state is a $1^+$. As the isovector strength is ramped up the $0^+$ state emerges as the ground state.  Thus in the presence of non-degenerate single-particle levels it is the interplay of isovector and isoscalar pairing that produces the ground state spin in $N=Z$ nuclei, a conclusion that in fact carries through to heavier $N=Z$ nuclei as we will discuss in a bit more detail later.

Next we consider the corresponding results for $^{42}Ca$, the parent nucleus for GT decay. These results are shown in Figs. 3 and 4 for Models 1 and 2, respectively. We only present the levels of the ground state rotational band in these figures.

\begin{figure}[hbtp]
\centering
\includegraphics[width=1.0\textwidth]{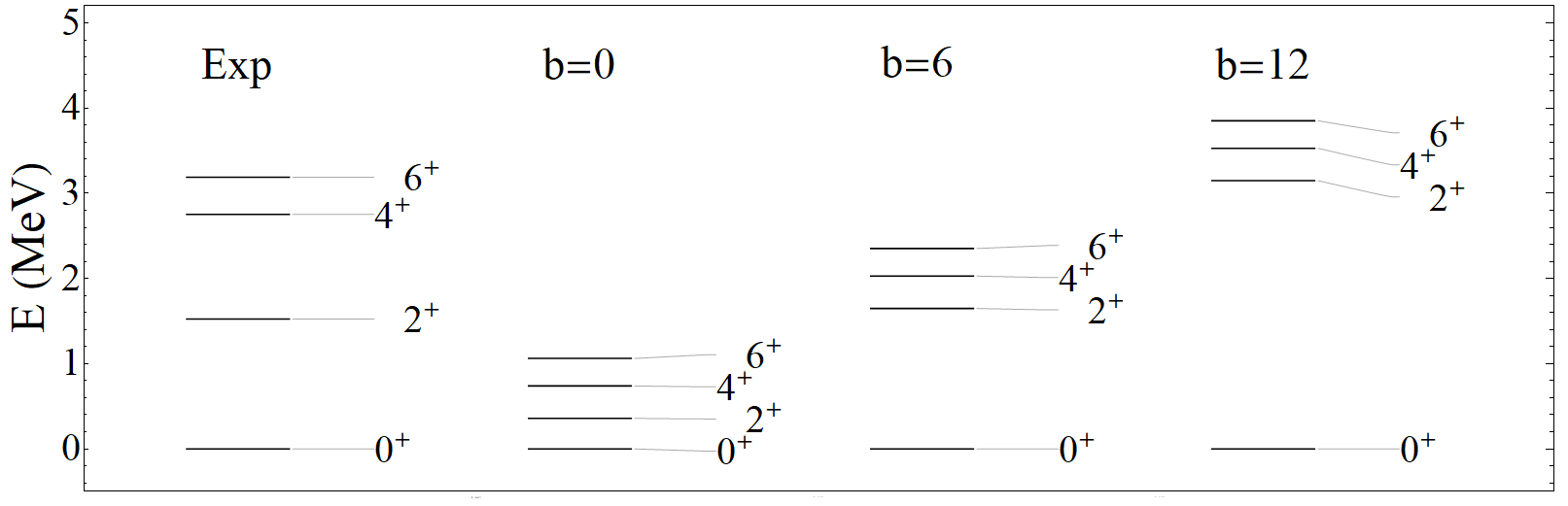}
\caption{Rotational band of $^{42}Ca$ from Model 1 as a function of the strength of the isovector pairing interaction $b$.}
\end{figure}

\begin{figure}[hbtp]
\centering
\includegraphics[width=1.0\textwidth]{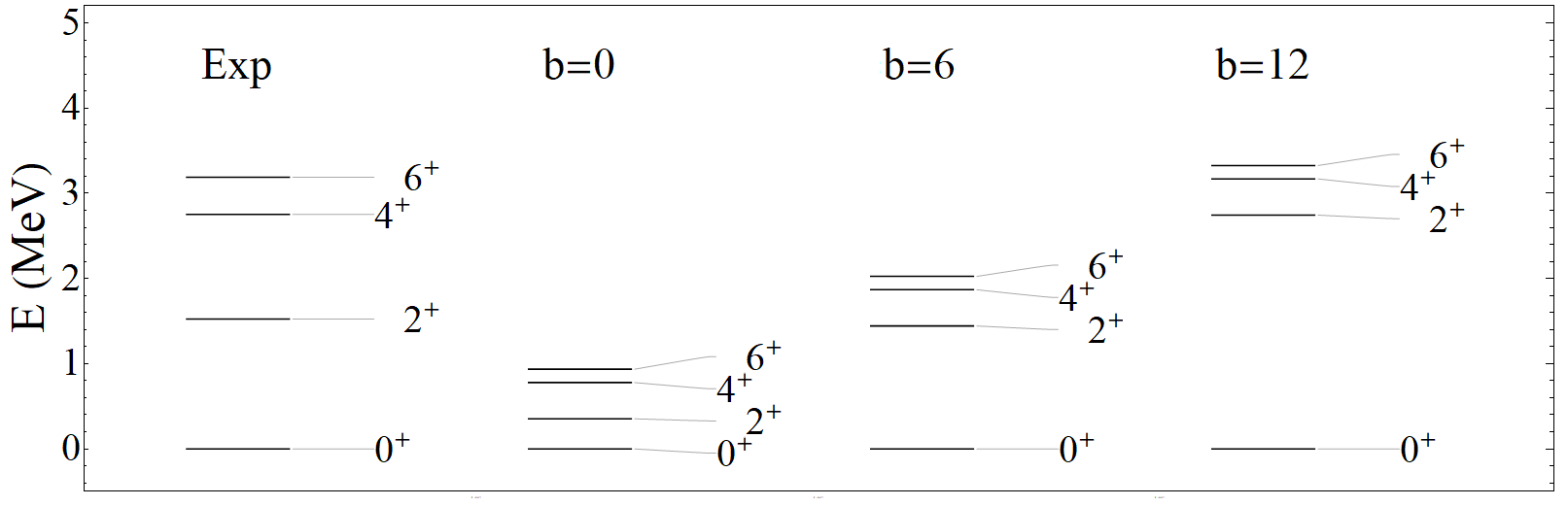}
\caption{Rotational band of $^{42}Ca$ from Model 2 as a function of the strength of the isovector pairing interaction $b$.}
\end{figure}

Since  there are only valence neutrons in $^{42}Ca$ the isoscalar pairing mode {\it cannot} contribute and thus it is omitted from the figures where we only present results as a function of the isovector strength parameter $b$. Note that for both models the spectrum does not exhibit a rotational pattern even for $b=0$. Clearly the presence of the single-particle energies erases the SU(3) rotational behavior of the purely quadrupole interaction, and in the presence of isovector pairing the rotational pattern is not recovered. In all cases, the $2^+$, $4^+$ and $6^+$ tend to group together, in marked contrast to the experimental data where only the $4^+$ and $6^+$ states are grouped.
This point notwithstanding good results seem to arise for both Model 1 and Model 2 when we choose $b=6$, in accord with the conclusion from the $^{42}Sc$ analysis. Other notable features that emerge are that: (1) as the isovector strength is reduced the overall spectrum is compressed, with the gap between the ground state and the first excited $2^+$ state becoming very small, and (2) the first excited $0^+$ state is invariably too high in energy.

It should be noted that this analysis is not the basis of our earlier remark that the isoscalar pairing term should be disregarded for $N \neq Z$ nuclei as it cannot play a role in a nucleus with only valence neutrons. When we look at the heavier systems to follow, the necessity of suppressing $a$ for $N \neq Z$ nuclei will become evident.

\subsubsection{A=44}

Next, we analyze the results obtained for nuclei with mass $ A = 44 $. As mentioned above, we use the same Hamiltonians as for $A=42$, except now we will {\it demonstrate} that isoscalar pairing must be removed when $N \neq Z$ so as to arrive at meaningful results.  As in the treatment of $A=42$, we will first discuss the spectra of the daughter ($^{44}Sc$) nucleus and then subsequently the parent ($^{44}Ca$) nucleus.

\begin{figure}[hbtp]
\centering
\includegraphics[width=1.0\textwidth]{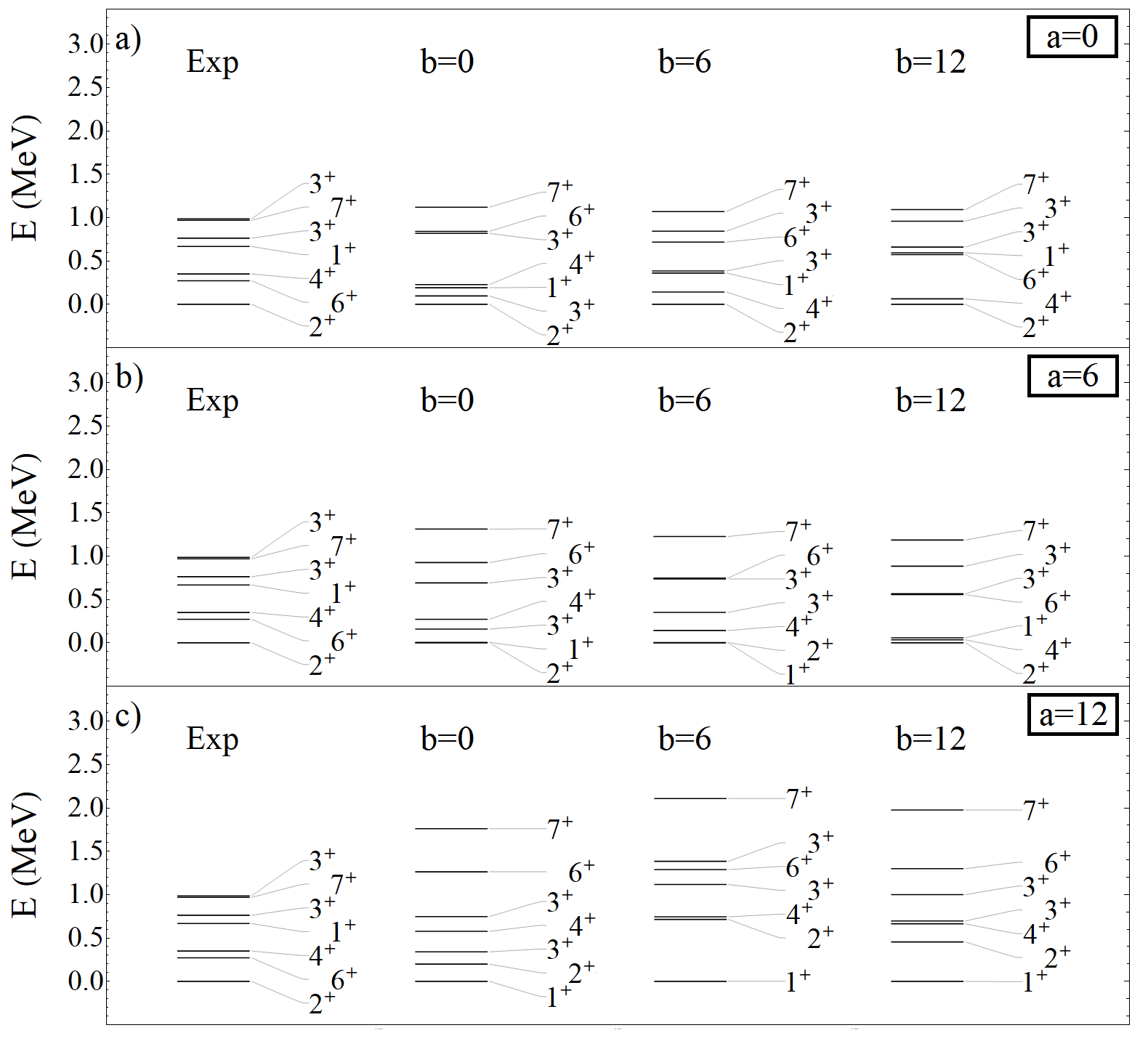}
\caption{Energy spectra of $^{44}Sc$ from Model 1 as a function of the isovector $b$ and isoscalar $a$  pairing strengths. Results are shown for $a,~b~=0,~6$ and $12$.}
\end{figure}

\begin{figure}[hbtp]
\centering
\includegraphics[width=1.0\textwidth]{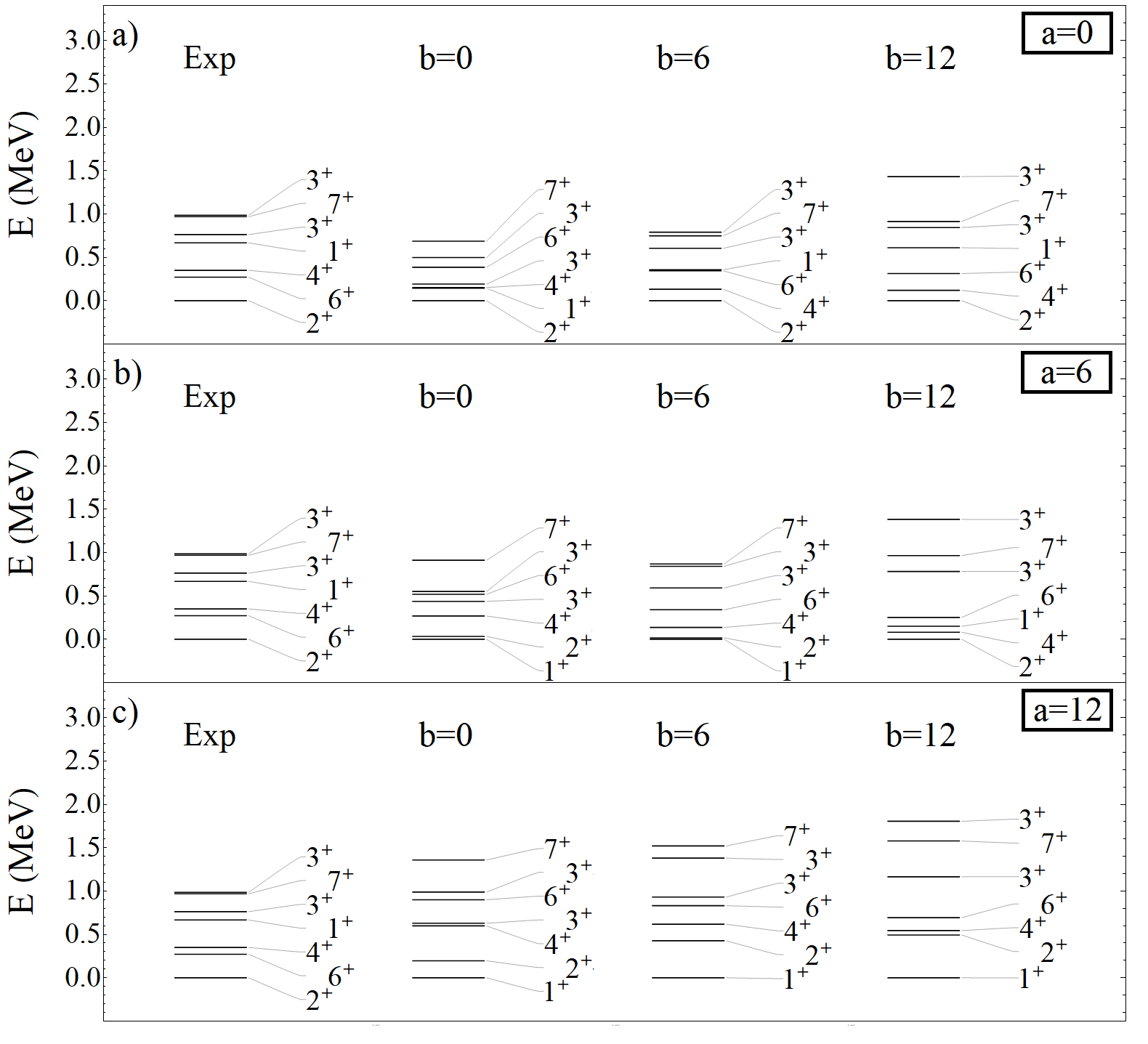}
\caption{Energy spectra of $^{44}Sc$ from Model 2 as a function of the isovector $b$ and isoscalar $a$  pairing strengths. Results are shown for $a,~b~=0,~6$, and $12$.}
\end{figure}

Results for the energy spectrum of $^{44}Sc$ of Model 1 are shown in Figure 5.  We note first that when we consider the equal pairing results $a=b$ we find that for the optimal value of the parameters $a=b=6$ the ground state spin (which is known to be $2^+$) is not reproduced. Rather we produce a ground state of angular momentum and parity $1 ^{+} $.
In contrast if we reduce the isoscalar pairing strength to $a=0$ we recover the correct ground state spin.  However, as for $^{42}Sc$ the $3^+$ and $7^+$ states are too high in energy even for $b=6$ and $a=0$. Reiterating what we said in our discussion of $A=42$ the simple model Hamiltonians we use are missing components that are needed to describe these states.

Similar results are seen with Model 2 as shown in Figure 6. Here too we see that  setting $a=b=6$ (as for the $N=Z$ nucleus) yields a $1^+$ ground state whereas $a=0$ and $b=6$ yields a $2^+$ ground state and a $1^+$ state at a reasonable excitation energy.

It is also worth commenting on the effect of changing only one of the pairing strengths at a time. For both models we see that the primary effect of reducing the isoscalar strength $a$ alone is to compact the set of odd-J states while leaving the relative spacing of the even-J states unaffected. In contrast when we reduce the isovector pairing strength $b$ alone the primary effect is to compact the set of even-J states while leaving the relative energies of the odd-J states unaffected.

Next we turn to the results for the parent nucleus $^{44}Ca$ and focus on the states of the energy spectra only. Since isoscalar pairing cannot contribute in nuclei with only valence neutrons we show in Figs. 7 and 8 the effect of isovector pairing only for Models 1 and 2, respectively. The features observed as a function of the isovector strength parameter are similar to those seen in $^{42}Ca$: the optimal energy spread arises for $b=6$, the spectrum is compressed as the isovector strength is decreased with the gap between the ground state and the first excited $2^+$ state becoming very small, and the first excited $0^+$ state remains too high in energy for all isovector strengths.

\begin{figure}[hbtp]
\centering
\includegraphics[width=1.0\textwidth]{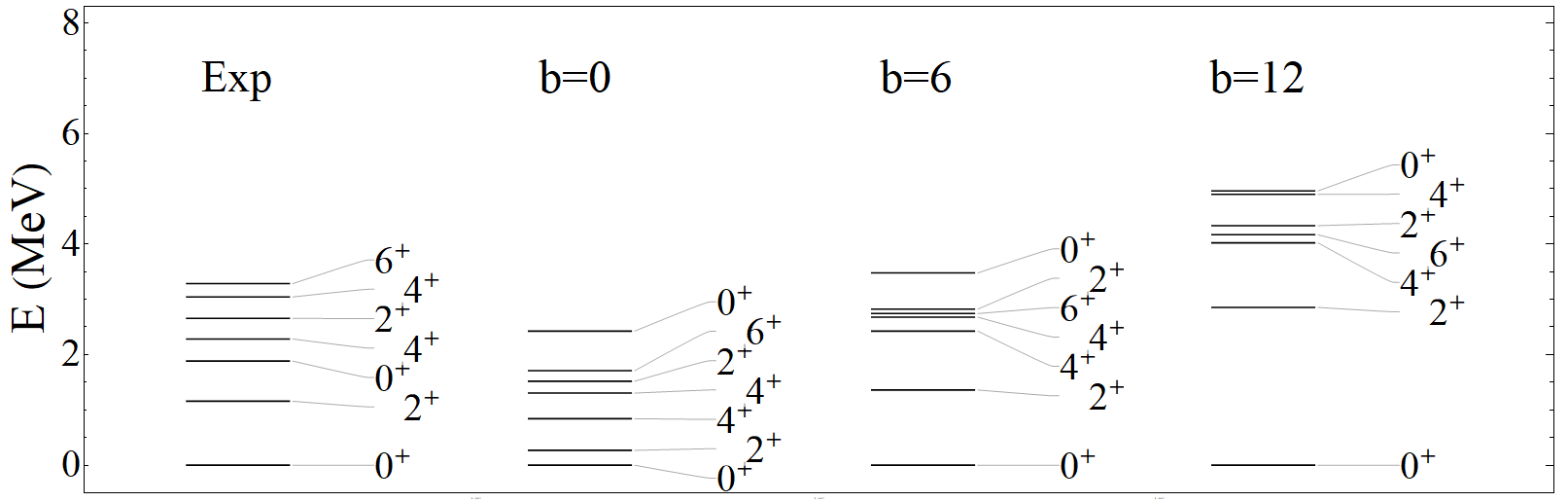}
\caption{ Energy spectra of $^{44}Ca$ from Model 1 as a function of the isovector pairing strength b.}
\end{figure}

\begin{figure}[hbtp]
\centering
\includegraphics[width=1.0\textwidth]{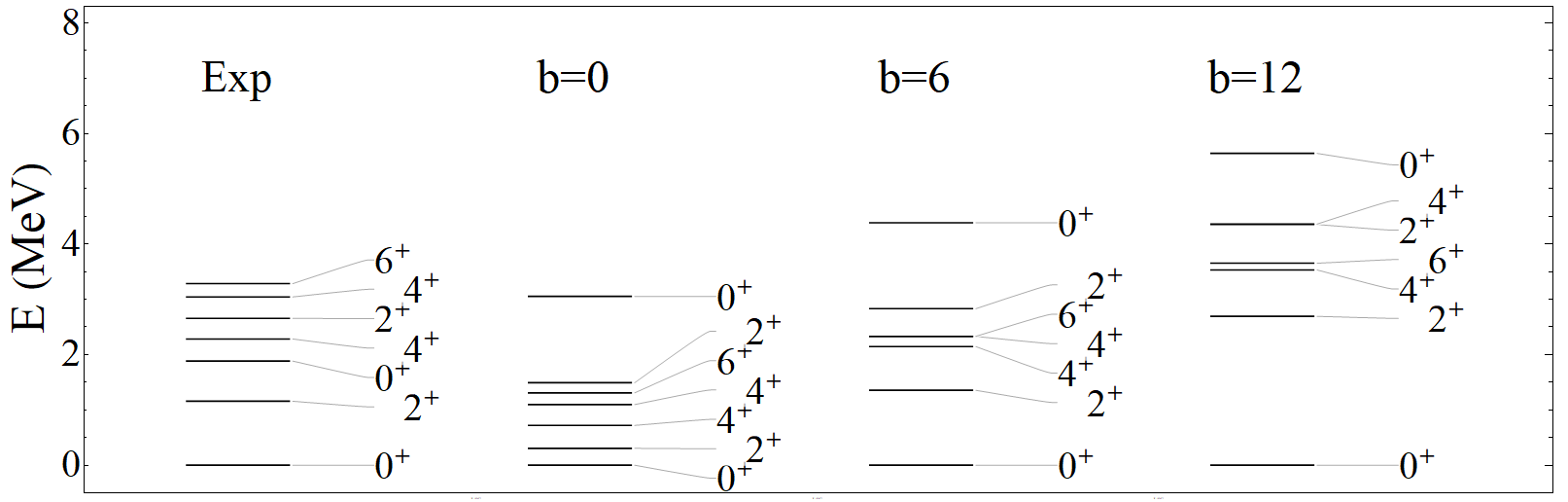}
\caption{ Energy spectra of $^{44}Ca$ from Model 2 as a function of the isovector pairing strength b.}
\end{figure}

\begin{figure}[hbtp]
\centering
\includegraphics[width=1.0\textwidth]{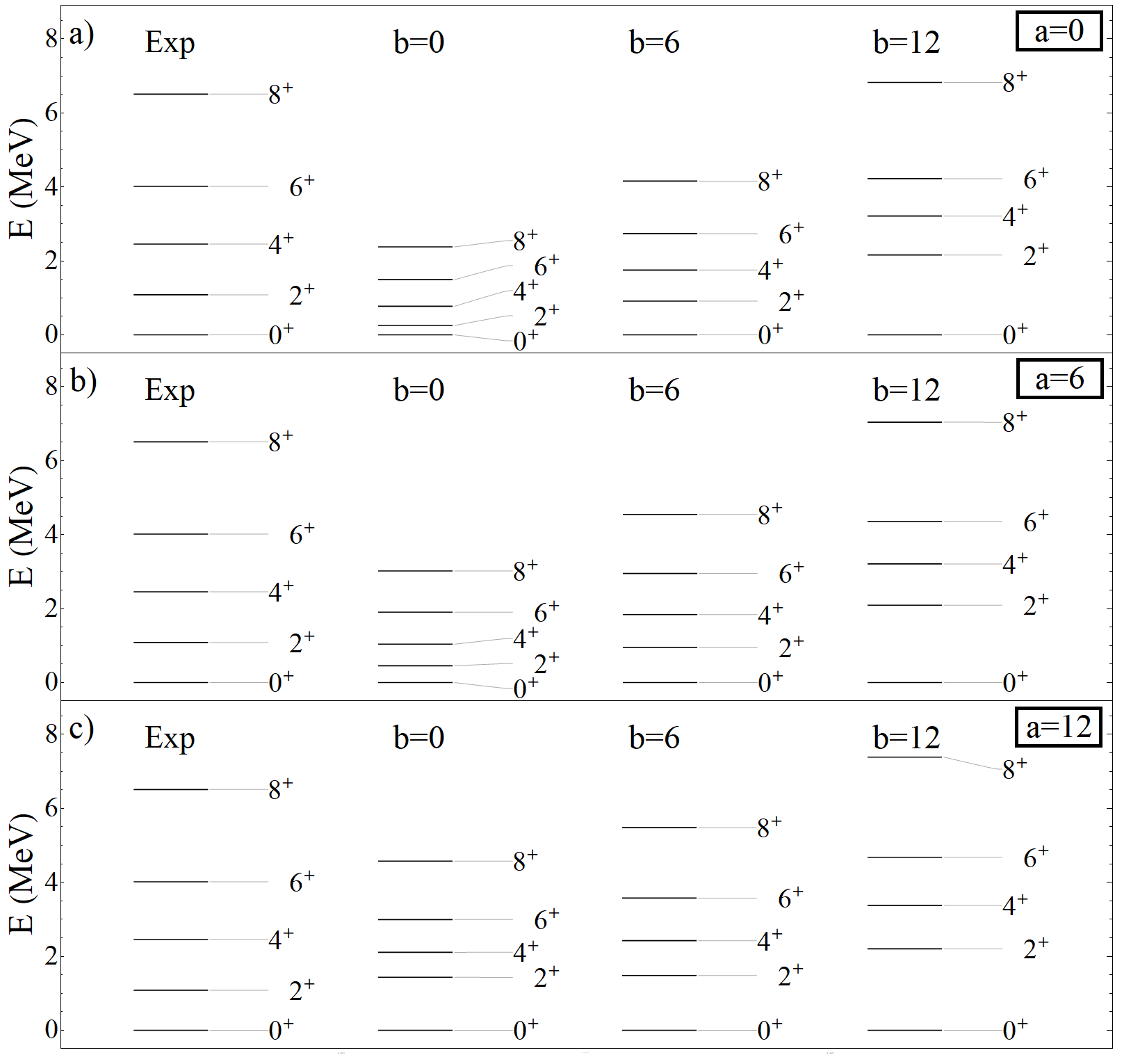}
\caption{Rotational band of $^{44}Ti$ from Model 1 as a function of the isovector $b$ and isoscalar $a$  pairing strengths. Results are shown for $a,~b~=0,~6$, and $12$.}
\end{figure}

\begin{figure}[hbtp]
\centering
\includegraphics[width=1.0\textwidth]{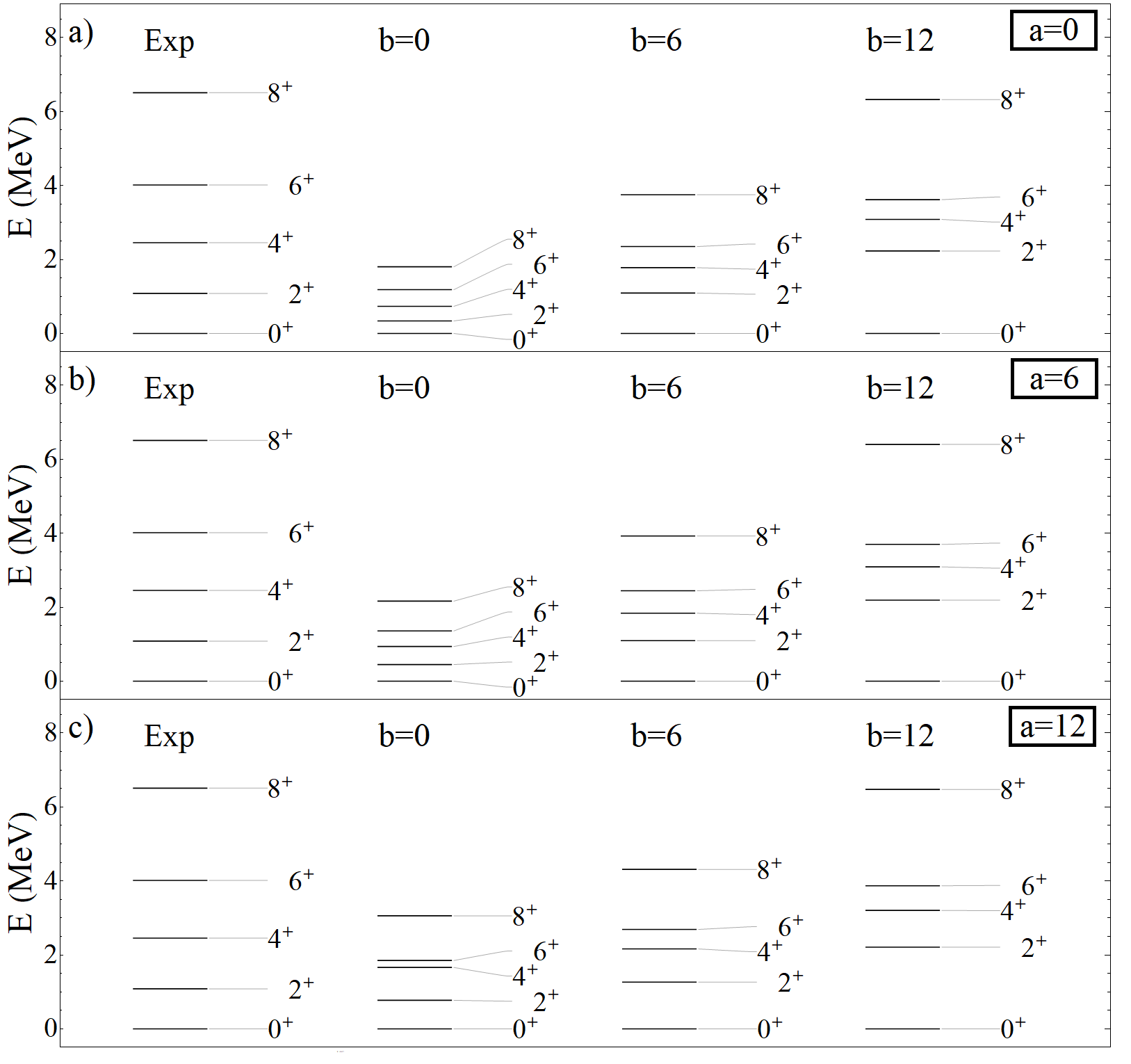}
\caption{Rotational band of $^{44}Ti$ from Model 2 as a function of the isovector $b$ and isoscalar $a$  pairing strengths. Results are shown for $a,~b~=0,~6$, and $12$.}
\end{figure}

Finally we consider the nucleus $^{44}Ti$ with 2 valence neutrons and 2 valence protons. Though not a part of the GT decay, it is worth looking at it nevertheless to study the effect of the different pairing modes on its rotational properties. As a reminder without pairing and single-particle effects the quadrupole-quadrupole interaction would produce a pure rotor with SU(3) symmetry. Figs. 9 and 10 show the influence of the two pairing modes on the rotational band properties for Models 1 and 2, respectively. The first point to note is that in the complete absence of pairing, i.e. $a=b=0$, the rotational band is already perturbed, through the presence of the single-particle field. Both isovector and isocalar pairing spread the spectrum, in better agreement with experiment, but do not restore the pure rotational character.

\subsubsection{A=46}

Next we consider the energy spectra of the $A=46$ nuclei involved in the GT decay $^{46}Ti~\rightarrow ~^{46}V$.

The results for $^{46}V$ are presented in Fig. 11 and 12 for Models 1 and 2, respectively.
For both models the results exhibit the basic features discussed earlier for lighter nuclei.  First, when the intensity of both pairings decrease simultaneously, the spectrum energies are reduced.
Next, when we reduce the isoscalar pairing alone (note this is an $N=Z$ nucleus so that isoscalar pairing is relevant), there is no appreciable effect on the states with even angular momentum, whereas those with  odd angular momentum gradually compress. Finally, when decreasing the isovector strength alone, a compression of the overall spectrum occurs and the ground state gradually changes from $ 0^{+} $ to $ 1^{+} $.

\begin{figure}[hbtp]
\centering
\includegraphics[width=1.0\textwidth]{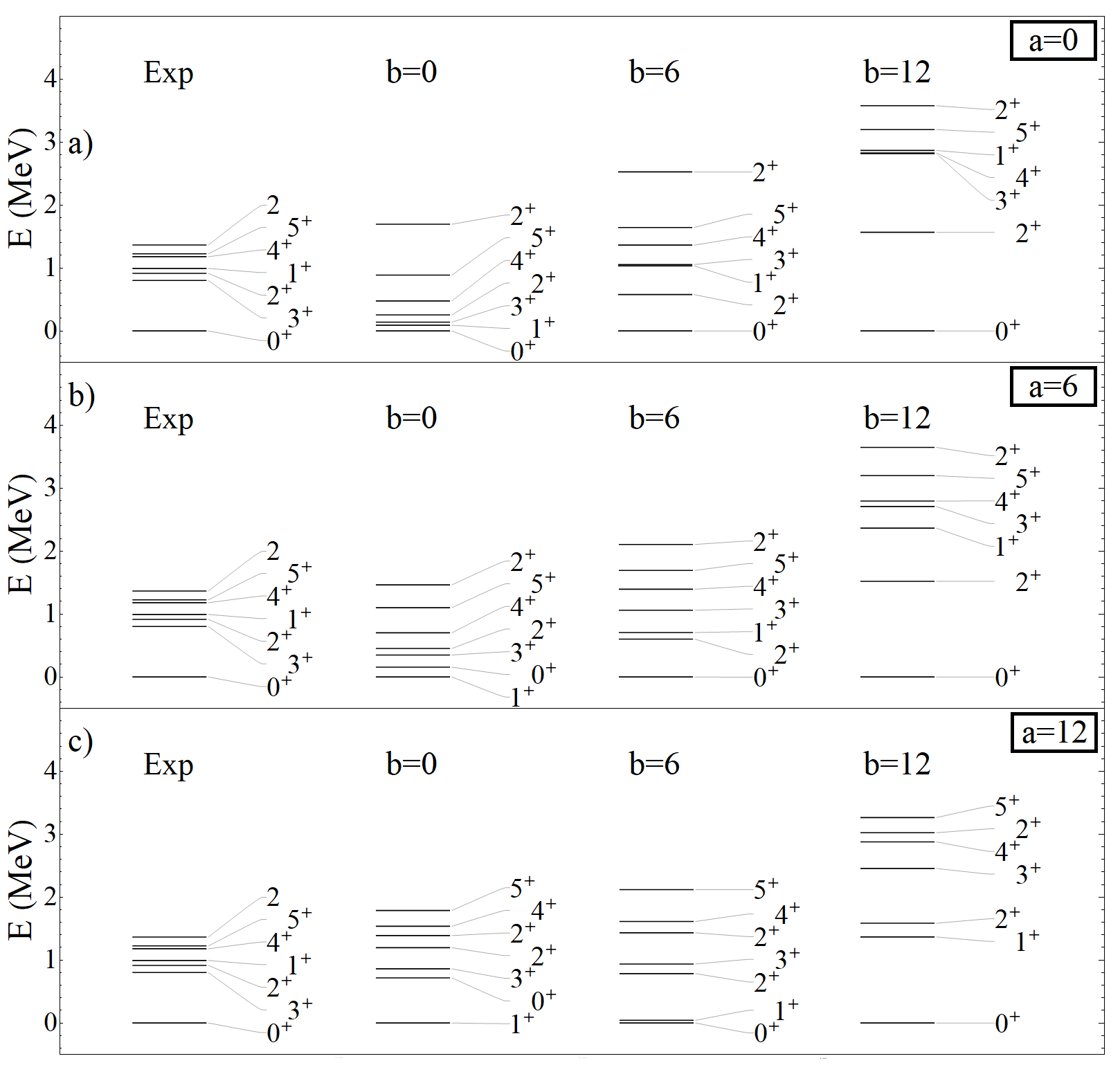}
\caption{Energy spectra of $^{46}V$ from Model 1 as a function of the isovector $b$ and isoscalar $a$  pairing strengths. Results are shown for $a,~b~=0,~6$, and $12$.}
\end{figure}

\begin{figure}[hbtp]
\centering
\includegraphics[width=1.0\textwidth]{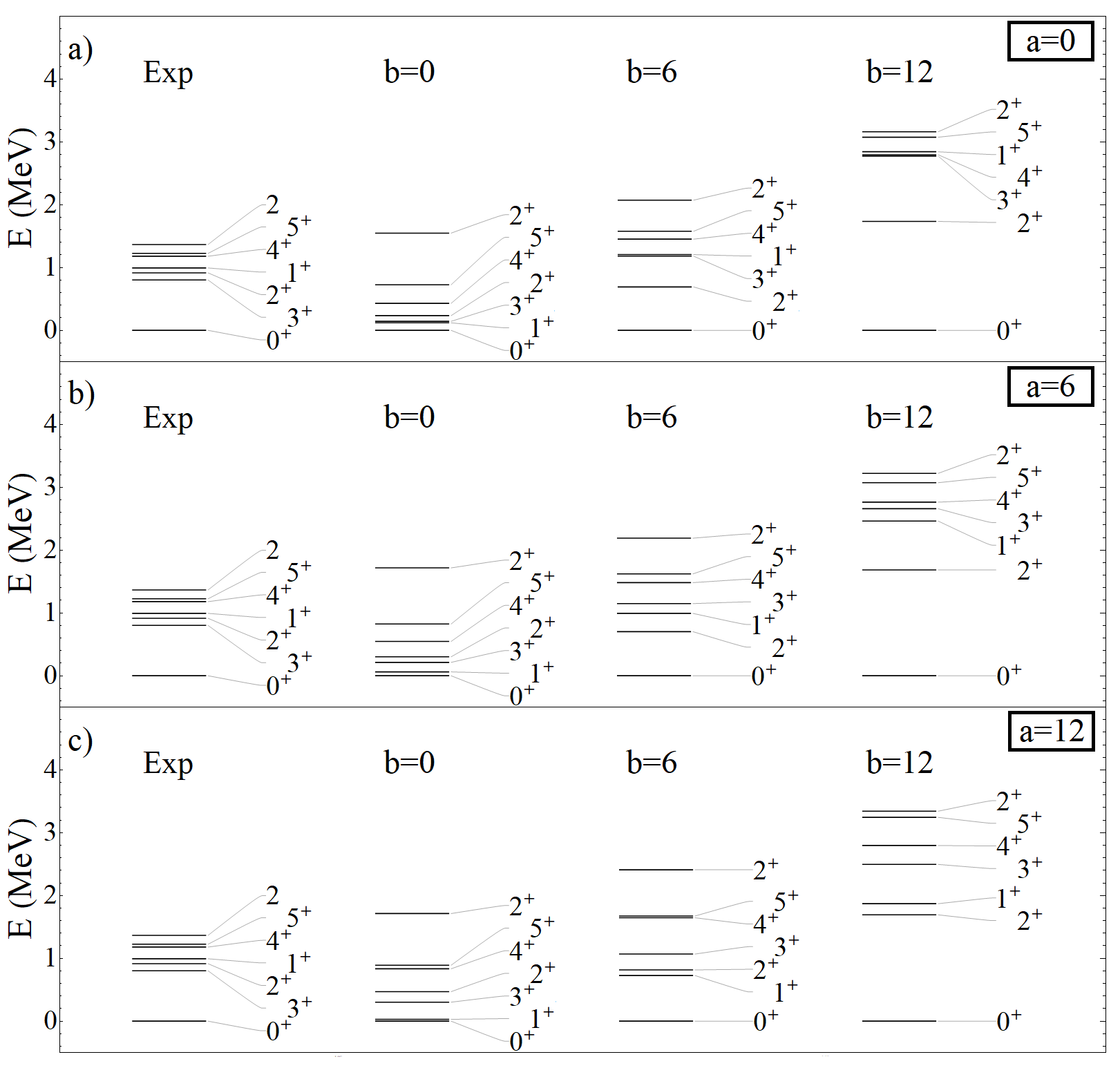}
\caption{Energy spectra of $^{46}V$ from Model 2 as a function of the isovector $b$ and isoscalar $a$  pairing strengths. Results are shown for $a,~b~=0,~6$, and $12$.}
\end{figure}

The properties of $^{46}V$  under Model 1 were discussed earlier in \cite{Stu} and showed some interesting features that we briefly repeat here. Were there no spin-orbit term, i.e. $\alpha=0$, the results for any value of $a=b$ would yield a degenerate $0^+$ and $1^+$ ground state. As $\alpha$ is increased the $1^+$ state gradually increases in energy. As shown in that earlier work, for $a=b=12$ and $\chi=-0.05$ the calculated splitting was in fairly good agreement with the experimental value. In our case for $a=b=6$ and $\chi=-0.065$ good agreement is likewise achieved.

Next we turn our discussion to $^{46}Ti$ for which the relevant results for the rotational band are shown in Figs. 13 and 14 for Models 1 and 2, respectively. Though $^{46}Ti$ has a neutron excess we consider the effect of varying the isoscalar pairing strength $a$ to confirm that even in a nucleus with several neutrons and protons its effect is minimal. And indeed from Figs. 13 and 14 we see that the dependence on the isoscalar pairing strength $a$ is negligible in the lower part of the spectrum and weak, but observable, in the upper part.

\begin{figure}[hbtp]
\centering
\includegraphics[width=1.0\textwidth]{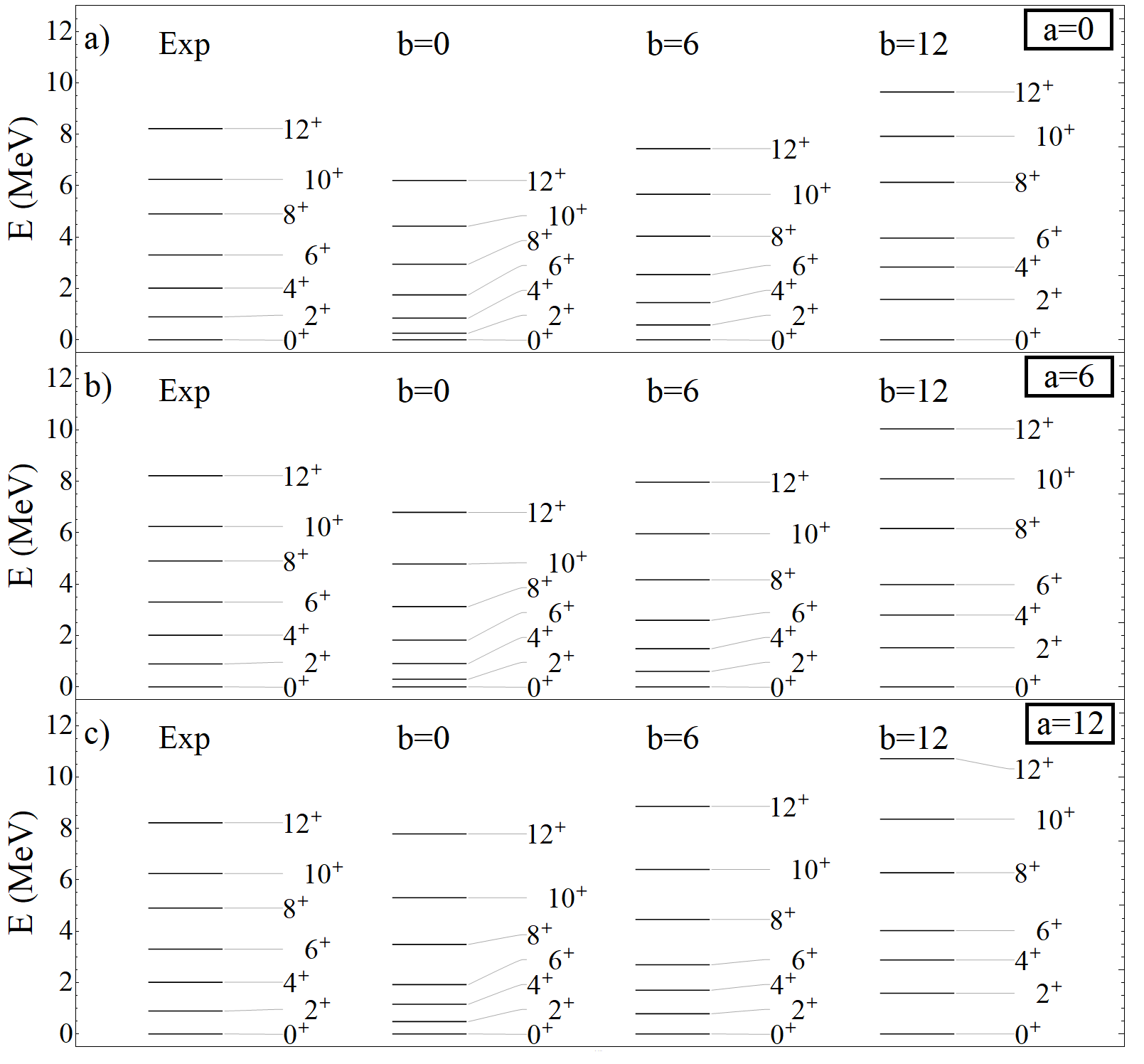}
\caption{Rotational band of $^{46}Ti$ from Model 1 as a function of the isovector $b$ and isoscalar $a$  pairing strengths. Results are shown for $a,~b~=0,~6$, and $12$es.}
\end{figure}

\begin{figure}[hbtp]
\centering
\includegraphics[width=1.0\textwidth]{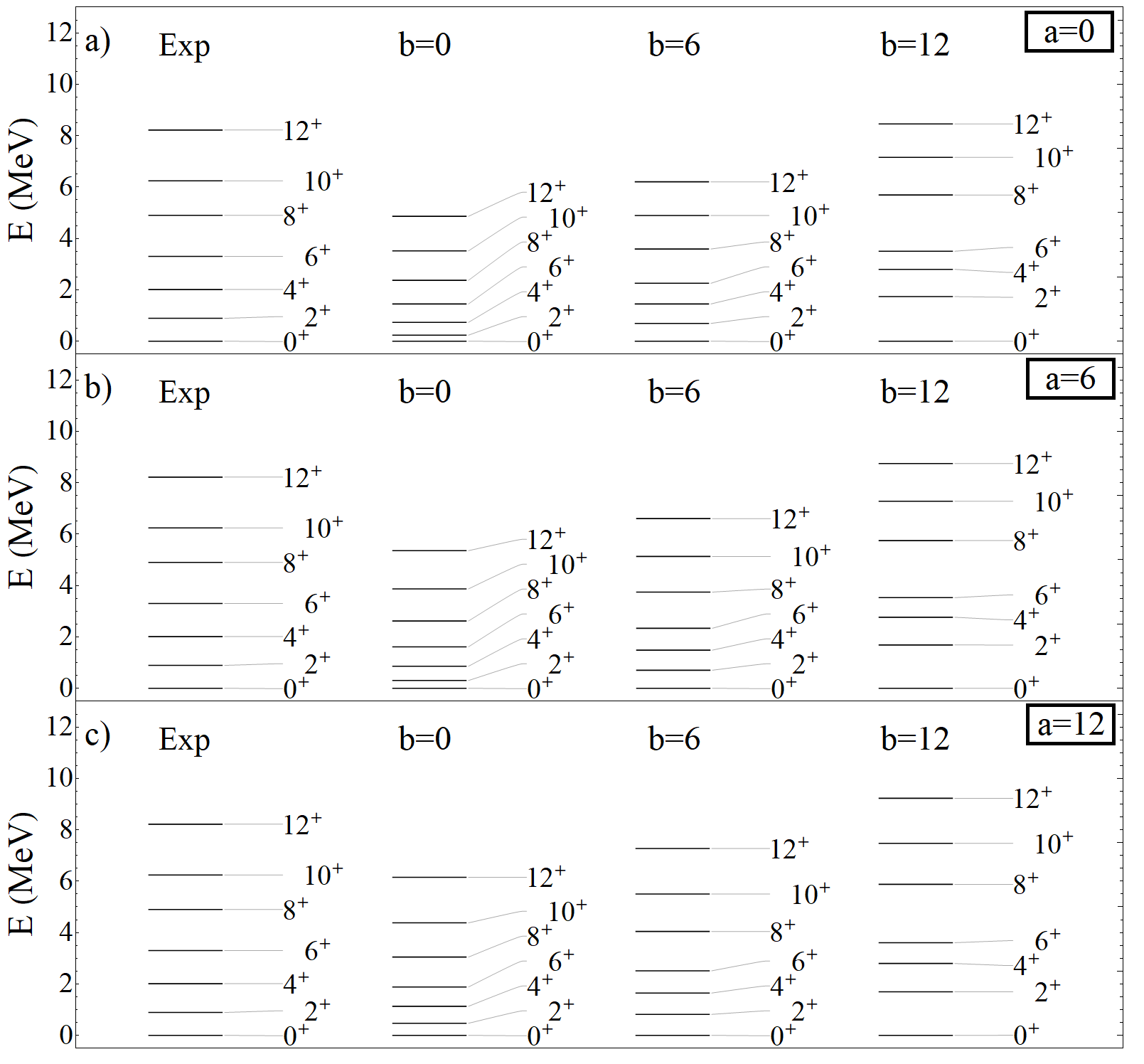}
\caption{Rotational band of $^{46}Ti$ from Model 2 as a function of the isovector $b$ and isoscalar $a$  pairing strengths. Results are shown for $a,~b~=0, 6$, and $12$.}
\end{figure}

In contrast, as can be seen from Figs. 13 and 14 the effect of isovector pairing is substantially more pronounced. Reducing its strength from the optimal value causes the spectra to be compacted, whereas increasing it leads to expansion of the spectrum.

Finally, as is evident from the data the rotational band of $ ^{46}Ti $ is not very rotational. Our results suggest that the optimal Hamiltonian of Model 1 ($a=b=6$ in Fig. 13) gives a good overall reproduction of the experimental data, and indeed substantially better than Model 2 ($a=b=6$ in Fig. 14).

\subsubsection{A=48}

Lastly we treat nuclei with $A=48$. We first discuss those that participate in the GT decay $^{48}Ti$ $\rightarrow$ $^{48}V$ and afterwards discuss $^{48}Cr$.

The results for $^{48}V$ are shown in Figs. 15 and 16 for Models 1 and 2, respectively. We include $a \neq 0$ values even though this nucleus has a relatively large neutron excess, for reasons that will be made clearer soon.

\begin{figure}[hbtp]
\centering
\includegraphics[width=1.0\textwidth]{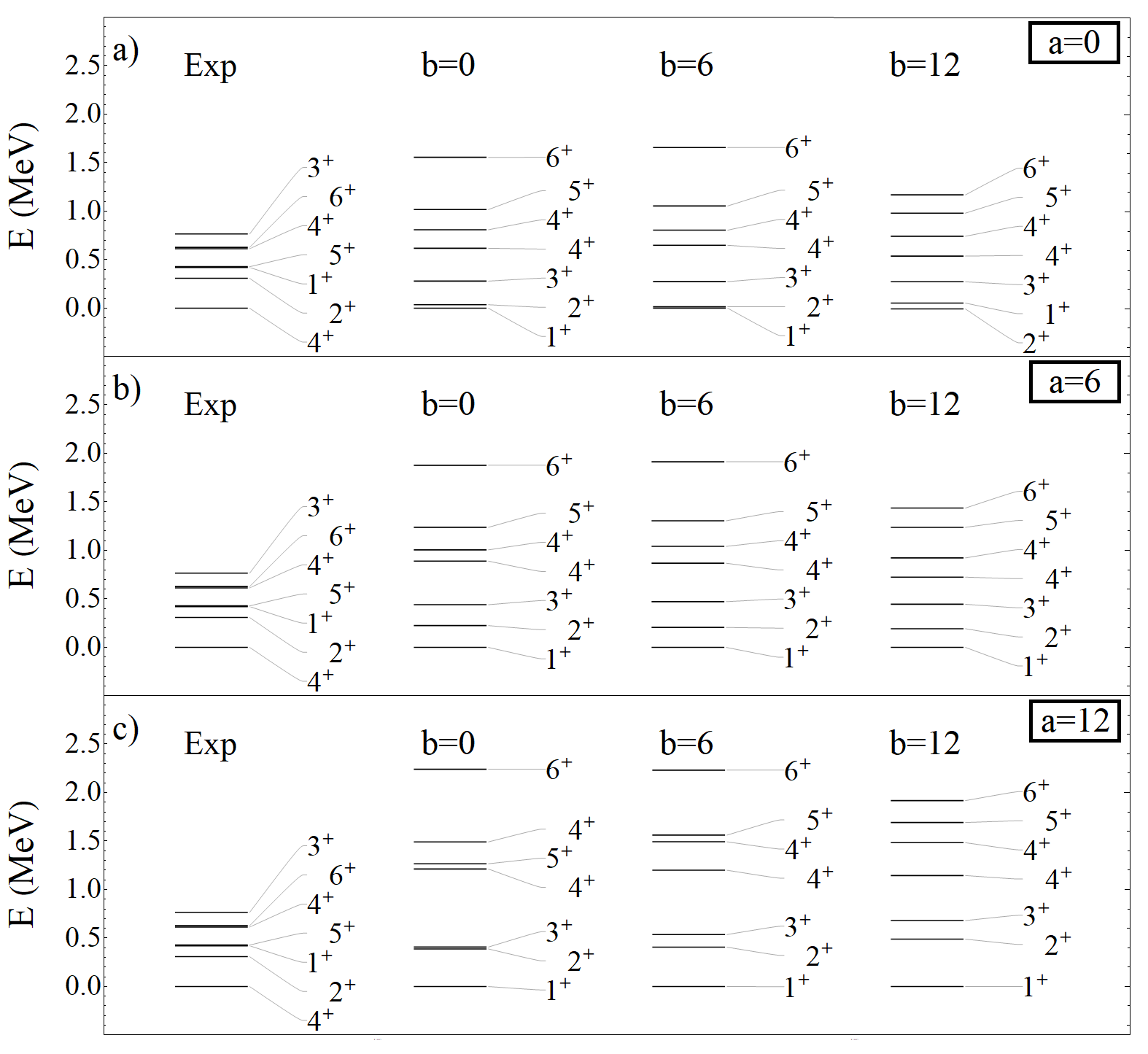}
\caption{ Energy spectra of $^{48}V$ from Model 1 as a function of the isovector $b$ and isoscalar $a$  pairing strengths. Results are shown for $a,~b~=0,~6$, and $12$.}
\end{figure}

\begin{figure}[hbtp]
\centering
\includegraphics[width=1.0\textwidth]{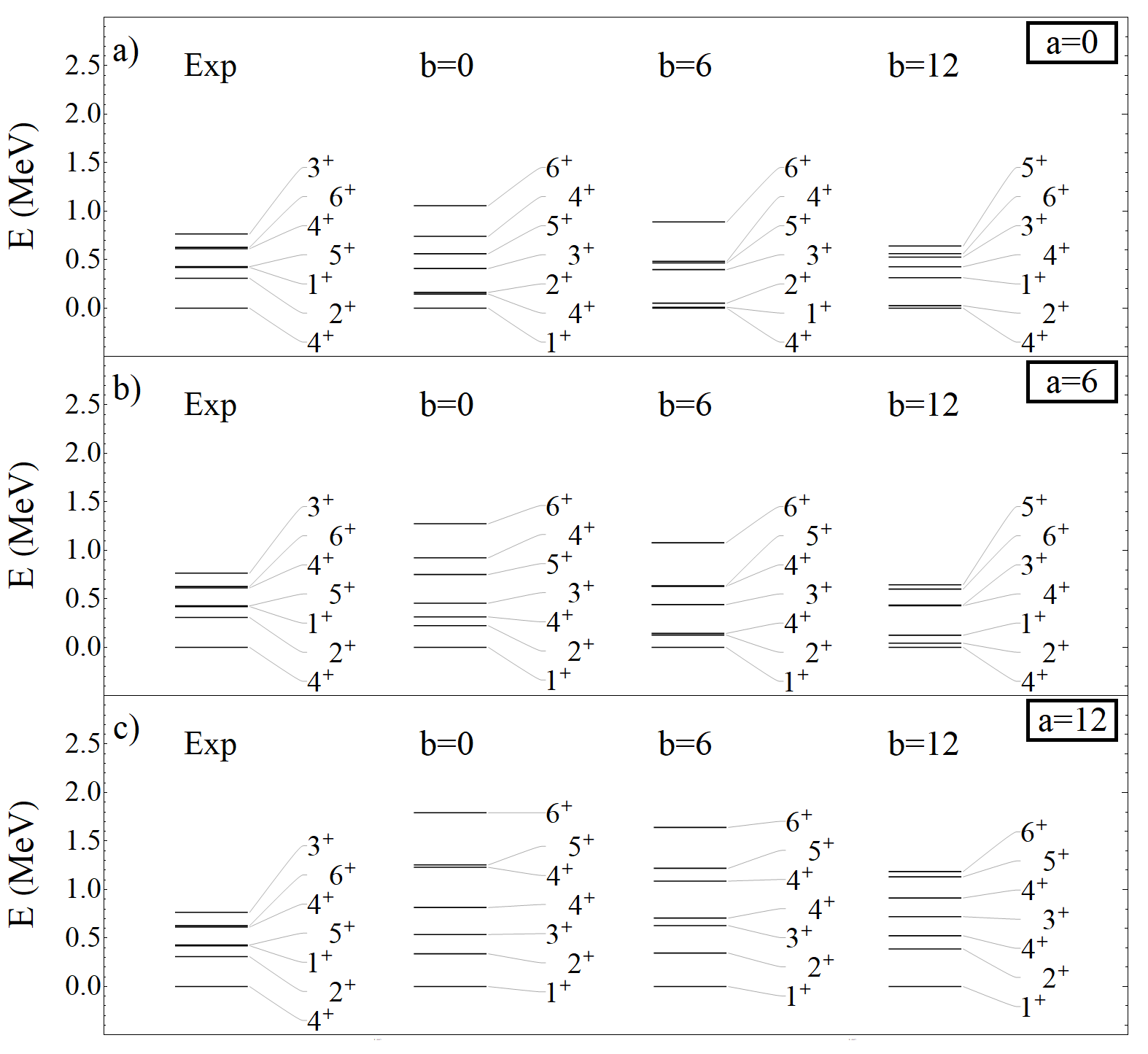}
\caption{Energy spectra of $^{48}V$ from Model 2 as a function of the isovector $b$ and isoscalar $a$  pairing strengths. Results are shown for $a,~b~=0,~6$, and $12$.}
\end{figure}

The first point to note is that the experimental ground state of this nucleus has spin and parity $4^+$. Model 1 is unable to reproduce this for {\it any} values of the two pairing parameters.

In contrast, Model 2 is able to reproduce the $4^+$ ground state for the optimal set of parameters $a=0$ and $b=6$. As $a$ is increased, however, it no longer produces the correct ground state spin and parity, confirming that here too it is critical to set $a=0$ to suppress erroneous features in the low-energy spectrum.

Results for the energy spectra $^{48}Ti$ are exhibited in Figs. 17 and 18 for Models 1 and 2, respectively. Model 1 with its optimal parameters $b=6$ and $a=0$ is able to achieve a good description of the low-energy spectrum. Model 2 with the same parameters is unable to reproduce the experimental results as well.

\begin{figure}[hbtp]
\centering
\includegraphics[width=1.0\textwidth]{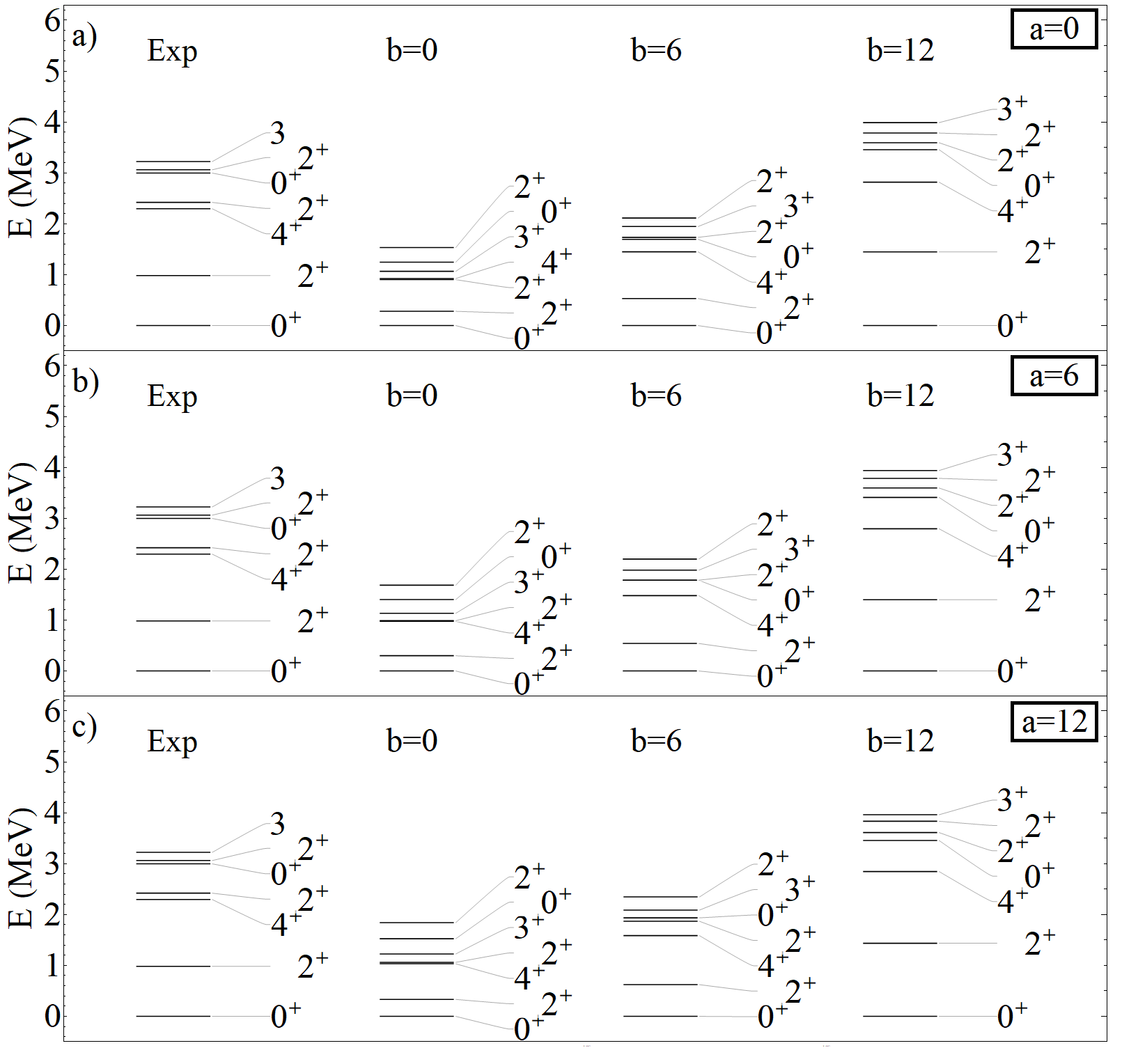}
\caption{Energy spectra of $^{48}Ti$ from Model 2 as a function of the isovector $b$ and isoscalar $a$  pairing strengths. Results are shown for $a,~b~=0,~6$, and $12$.}
\end{figure}

\begin{figure}[hbtp]
\centering
\includegraphics[width=1.0\textwidth]{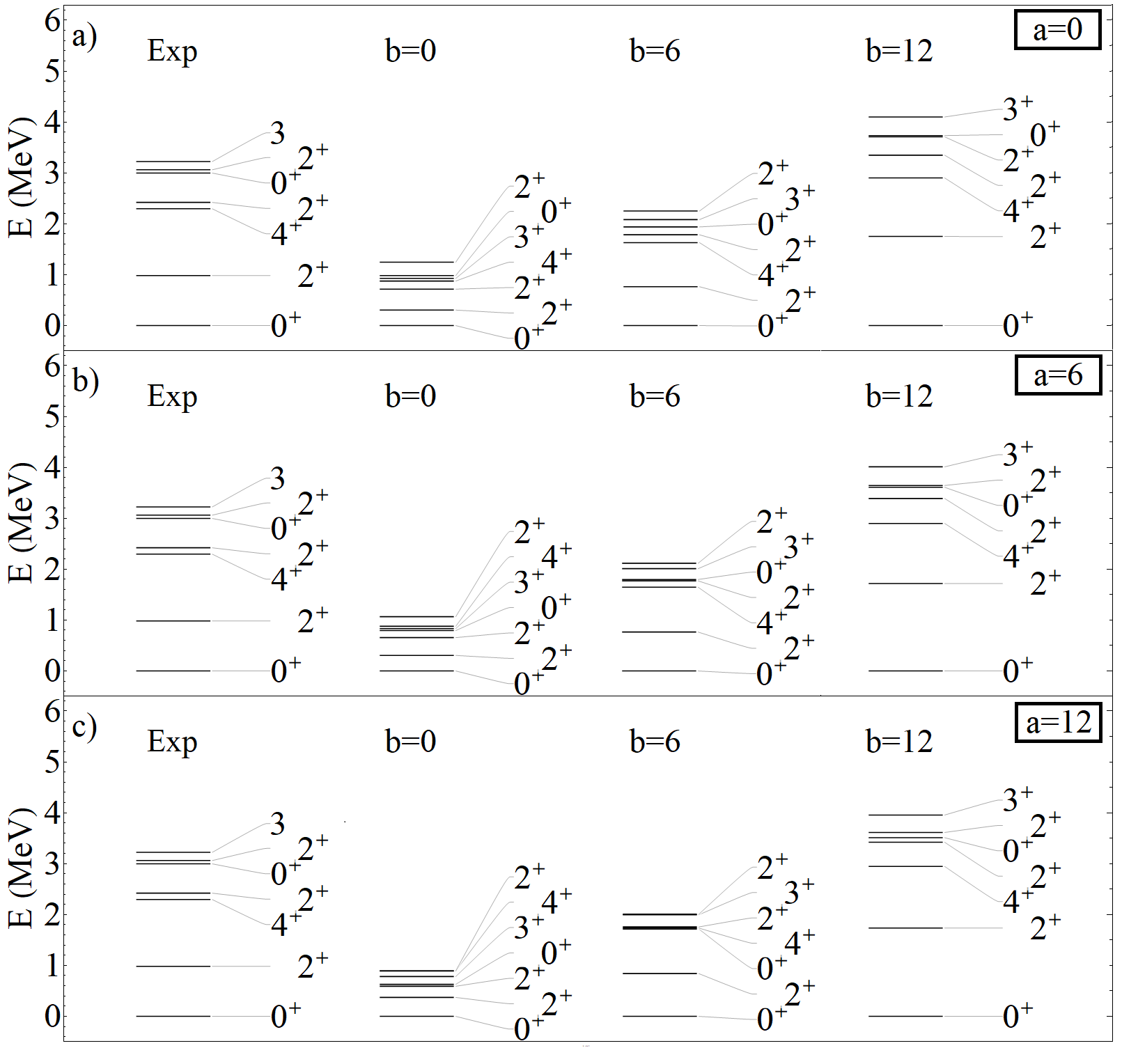}
\caption{Energy spectra of $^{48}Ti$ from Model 2 as a function of the isovector $b$ and isoscalar $a$  pairing strengths. Results are shown for $a,~b~=0,~6$, and $12$.}
\end{figure}

Finally we consider the $N=Z$ nucleus $^{48}Cr$. Here we show the results calculated for the states of the ground state rotational band, in Fig. 19 for Model 1 and Fig. 20 for Model 2.

\begin{figure}[hbtp]
\centering
\includegraphics[width=1.0\textwidth]{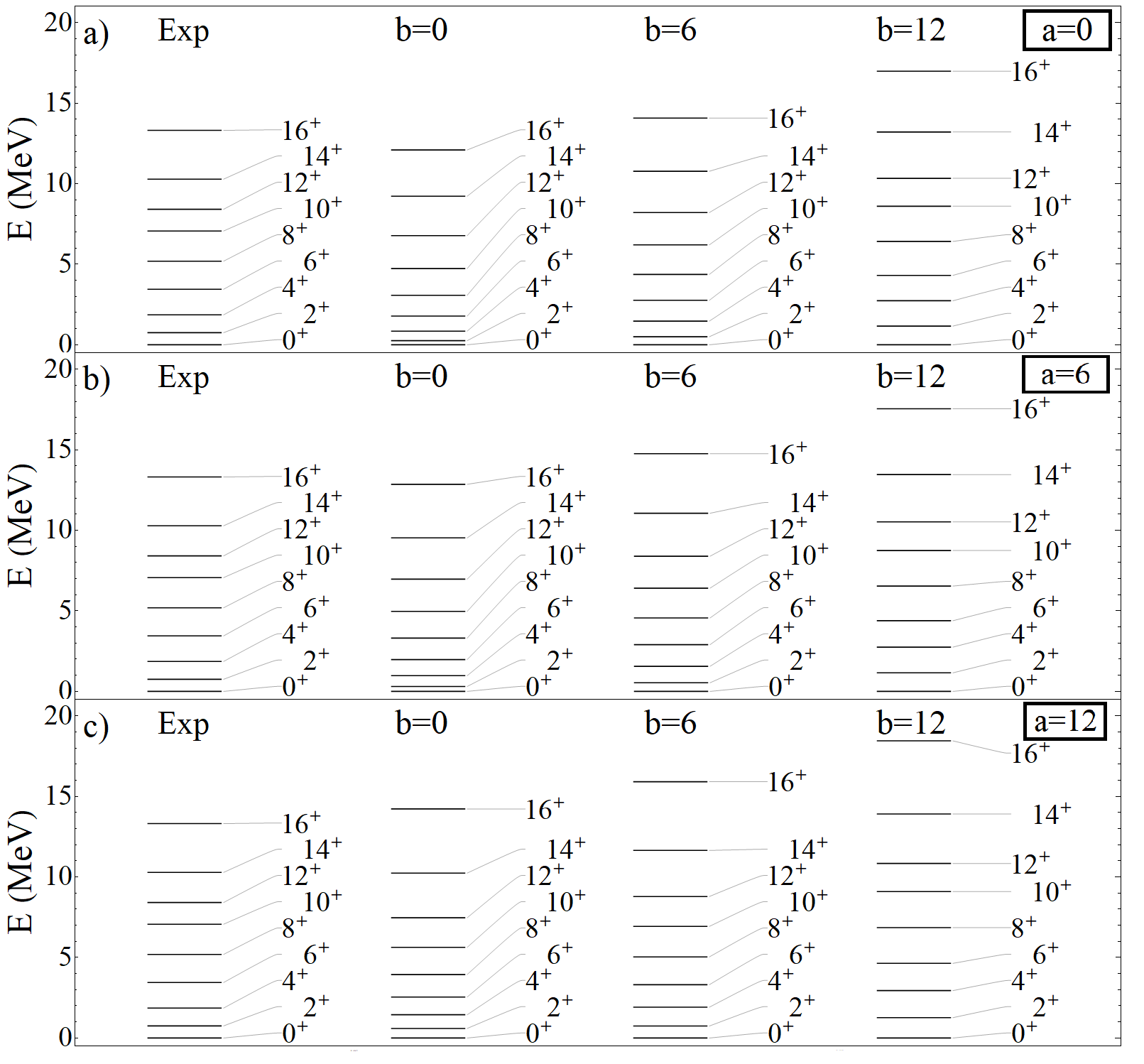}
\caption{Rotational band of $^{48}Cr$ from Model 1 as a function of the isovector $b$ and isoscalar $a$  pairing strengths. Results are shown for $a,~b~=0,~6$, and $12$.}
\end{figure}

\begin{figure}[hbtp]
\centering
\includegraphics[width=1.0\textwidth]{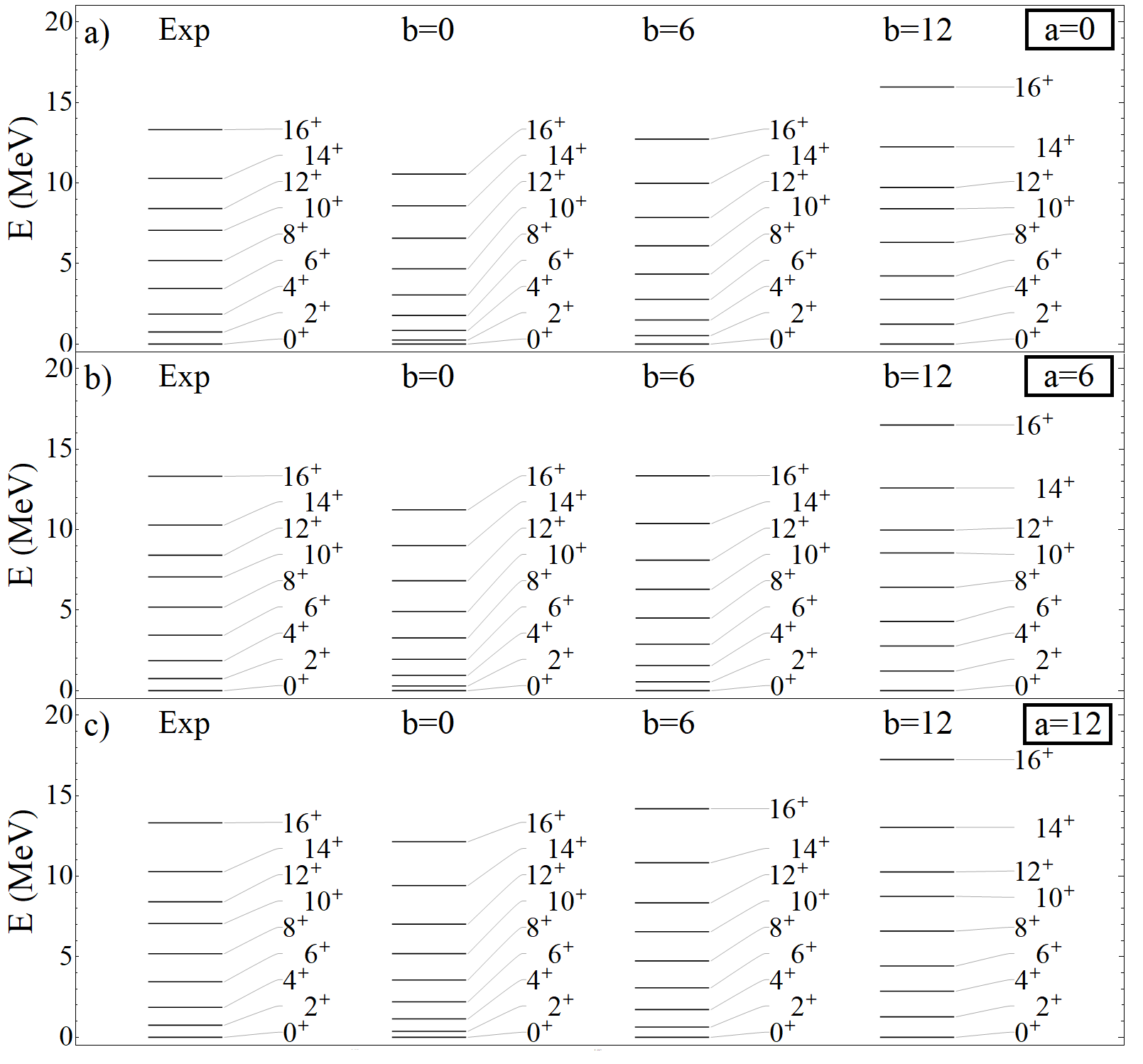}
\caption{Rotational band of $^{48}Cr$ from Model 2 as a function of the isovector $b$ and isoscalar $a$  pairing strengths. Results are shown for $a,~b~=0,~6$, and  $12$.}
\end{figure}

Both models with their optimal parametrizations are able to reproduce the general features of the ground state rotational band up to $J^{\pi}=16^+$. When we gradually suppress isovector pairing the spectra are compacted into a smaller energy range, especially for Model 2, and the overall agreement is lost.

\subsection{GT Transitions}

Below are the results corresponding to the calculation of GT transitions. The values of B(GT) are multiplied by the usual quenching factor $(0.74)^{2}$ \cite{quen,gysbers}.

\subsubsection{A=42}

We first consider the GT decay $^{42}Ca$ $\rightarrow$ $^{42}Sc$, for which the results are shown in Figs. 21 and 22 for Models 1 and 2, respectively. The results are shown as a function of the isovector pairing strength $b$, which applies to both the parent and daughter nucleus, {\it and} the isoscalar pairing strength $a$, which only contributes to the properties of the $^{42}Sc$ daughter nucleus.

\begin{figure}[hbtp]
\centering
\includegraphics[width=1.0\textwidth]{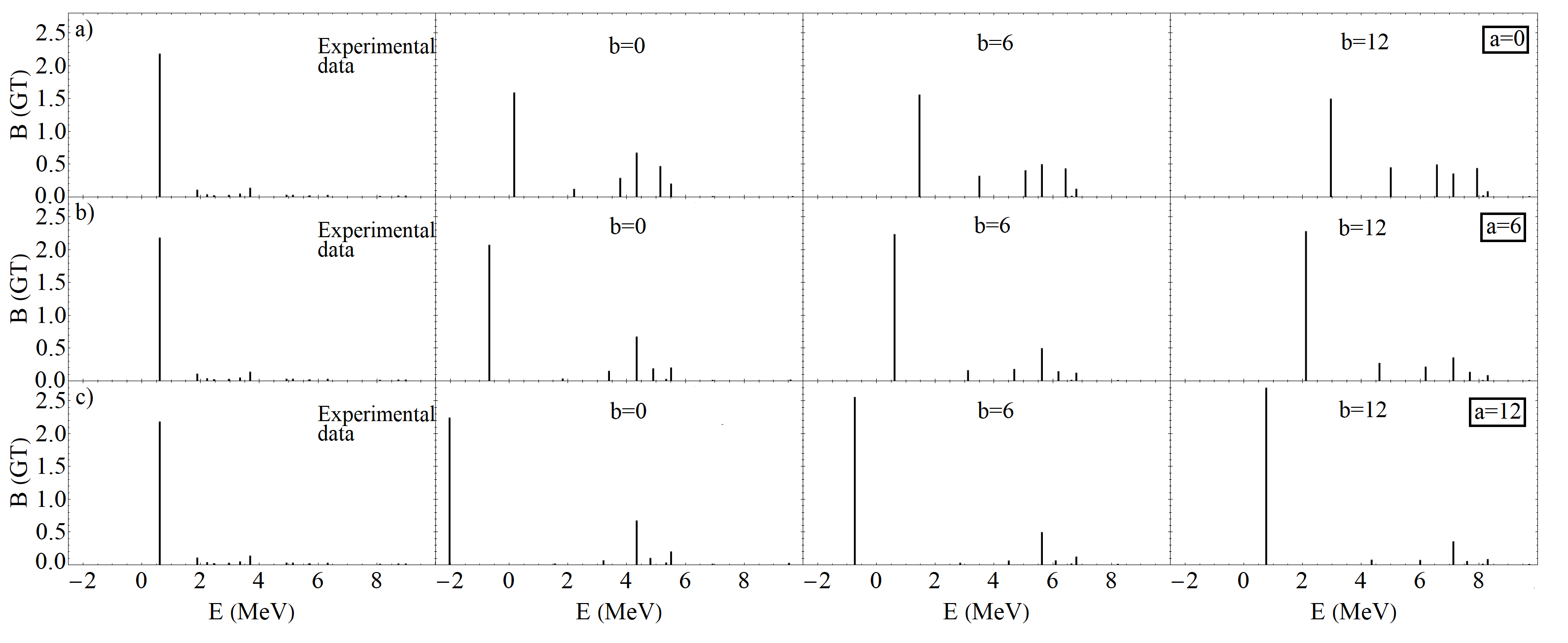}
\caption{Comparison of the experimental \cite{exp42} and the theoretical results of Model 1 for  B(GT) transition strengths for $^{42}Ca \rightarrow ^{42}Sc$ as a function of the isovector pairing strength $b$ and the isoscalar pairing strength $a$, which only acts in the daughter nucleus. }
\end{figure}

\begin{figure}[hbtp]
\centering
\includegraphics[width=1.0\textwidth]{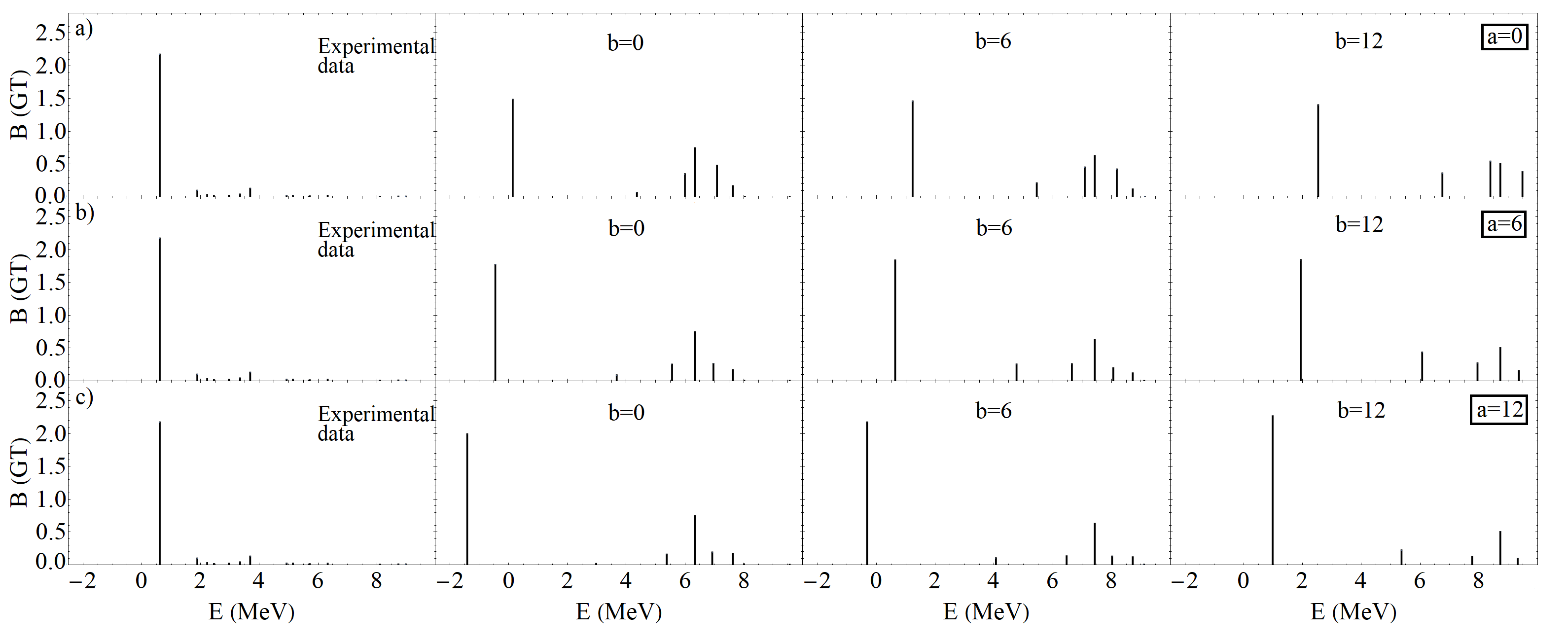}
\caption{Comparison of the experimental \cite{exp42} and theoretical results of Model 2 for  B(GT) transition strengths for $^{42}Ca \rightarrow ^{42}Sc$ as a function of the isovector pairing strength $b$ and the isoscalar pairing strength $a$, which only acts in the daughter nucleus. }
\end{figure}

Both models show very similar results. In both, the $a=b=6$ optimal pairing strengths for the $^{42}Sc$ daughter produce a single strong peak at almost the exact energy where it is seen experimentally. The strength is also close to that seen experimentally. Both also show small satellite peaks at higher energy with somewhat more strength than seen experimentally.

As $a$ is increased the strength to the lowest $1^+$ state increases, albeit slowly, while the energy of that state goes down in energy and eventually for $a=12$ becomes the ground state. As $b$ is increased, for a given $a$, the main peak moves up in energy, but with no noticeable change in its strength.

Reiterating, the same features appear for both our Model 1 and Model 2 Hamiltonians.

\subsubsection{A=44}

Now we analyze the results obtained for the nuclei of mass $ A = 44 $. In this case, both of the nuclei involved in the GT decay $^{44}Ca \rightarrow ^{44}Sc$ have a neutron excess, so that it is appropriate for our analysis to assume $a=0$ for both and to present the results as a function of the isovector pairing strength $b$ only. These results are presented in Figs. 23 and 24 for Models 1 and 2, respectively.

\begin{figure}[hbtp]
\centering
\includegraphics[width=1.0\textwidth]{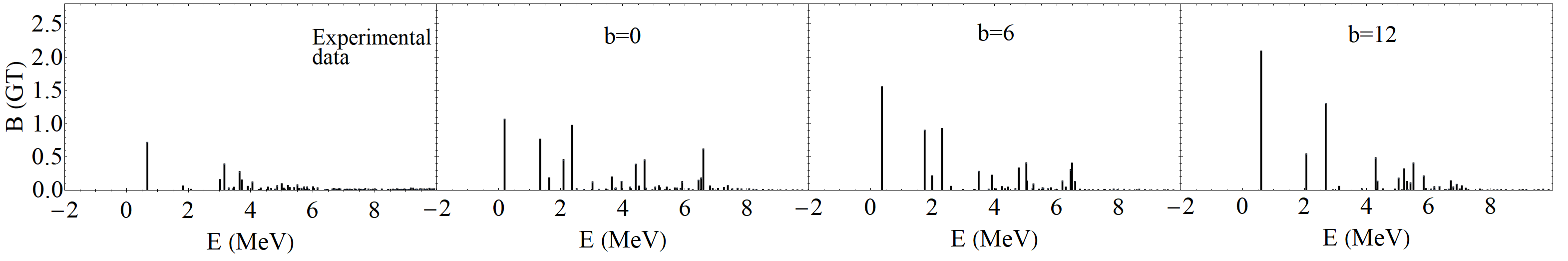}
\caption{Comparison of the experimental \cite{exp44} and theoretical results of Model 1 for  B(GT) transition strengths for $^{44}Ca \rightarrow ^{44}Sc$ as a function of the isovector pairing strength $b$. Since both nuclei involved have neutron excesses, the isoscalar pairing strength is set to $a=0$ for both.  }
\end{figure}

\begin{figure}[hbtp]
\centering
\includegraphics[width=1.0\textwidth]{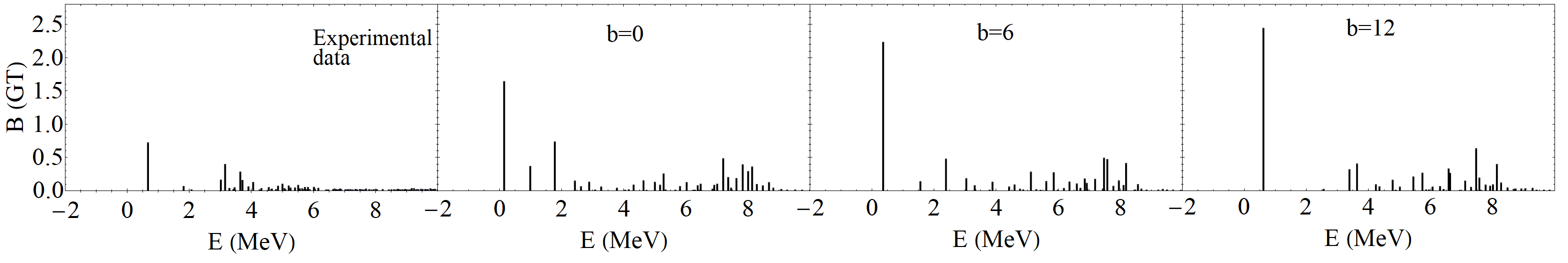}
\caption{Comparison of the experimental \cite{exp44} and theoretical results of Model 2 for  B(GT) transition strengths for $^{44}Ca \rightarrow ^{44}Sc$ as a function of the isovector pairing strength $b$. Since both nuclei involved have neutron excesses, the isoscalar pairing strength is set to $a=0$ for both.  }
\end{figure}

In this case, the results are similar for the two models, but with some notable differences.

For Model 1, the calculation produces several low-lying states with appreciable strength, as in the data. For $b=6$, the lowest peak is the strongest, and indeed stronger than in experiment. However, it is within a factor of roughly two of the strength of the next two peaks, in rough accord with experiment. As $b$ increases, the strength to the lowest state becomes increasingly larger, and the state moves up in energy.

In the case of Model 2, the enhancement in calculated strength to the lowest state is substantially more pronounced than for Model 1. Here too several satellite peaks appear at fairly low energies, but  relatively weaker than in Model 1. In the case of the optimal $b=6$ isovector strength the lowest excitation is roughly five times more strongly populated than the next few, in contrast with both experiment and Model 1 where the relative enhancement is roughly two. As for Model 1, the lowest peak moves up in energy as $b$ is increased and becomes progressively more dominant.

\subsubsection{ A=46}

Next, we analyze the GT results obtained for the nuclei with mass $ A = 46 $. While the parent nucleus involved in the GT decay $^{46}Ti \rightarrow ^{46}V$ has a neutron excess, the daughter nucleus does not. Thus we assume $a=0$ for the parent and present the results as a function of the $a$ value used in describing the daughter. In addition, the results are shown as a function of the isovector pairing strength $b$ used for both nuclei. These results are presented in Figs. 25 and 26 for Models 1 and 2, respectively.

\begin{figure}[hbtp]
\centering
\includegraphics[width=1.0\textwidth]{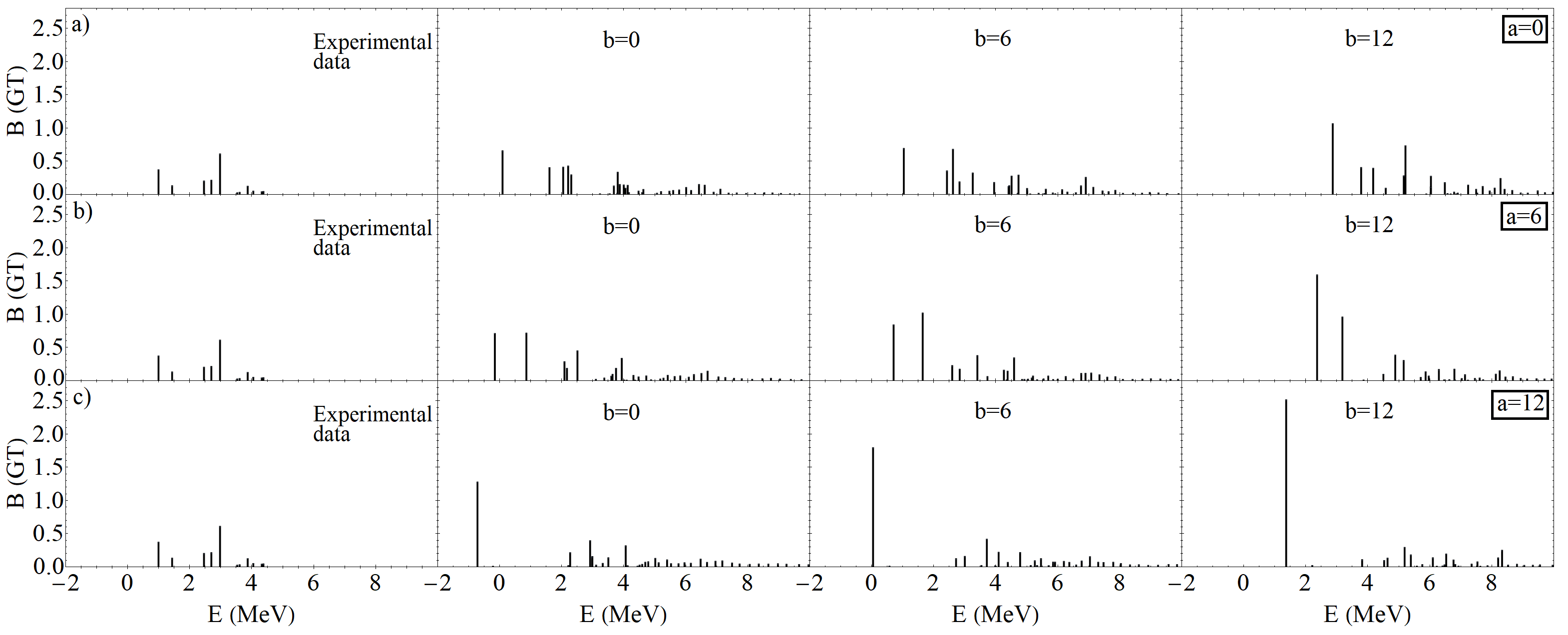}
\caption{Comparison of the experimental \cite{exp46} and the theoretical results of Model 1 for  B(GT) transition strengths for $^{46}Ti \rightarrow ^{46}V$ as a function of the isovector pairing strength $b$ and the isoscalar pairing strength $a$, which only acts in the daughter nucleus. }
\end{figure}

\begin{figure}[hbtp]
\centering
\includegraphics[width=1.0\textwidth]{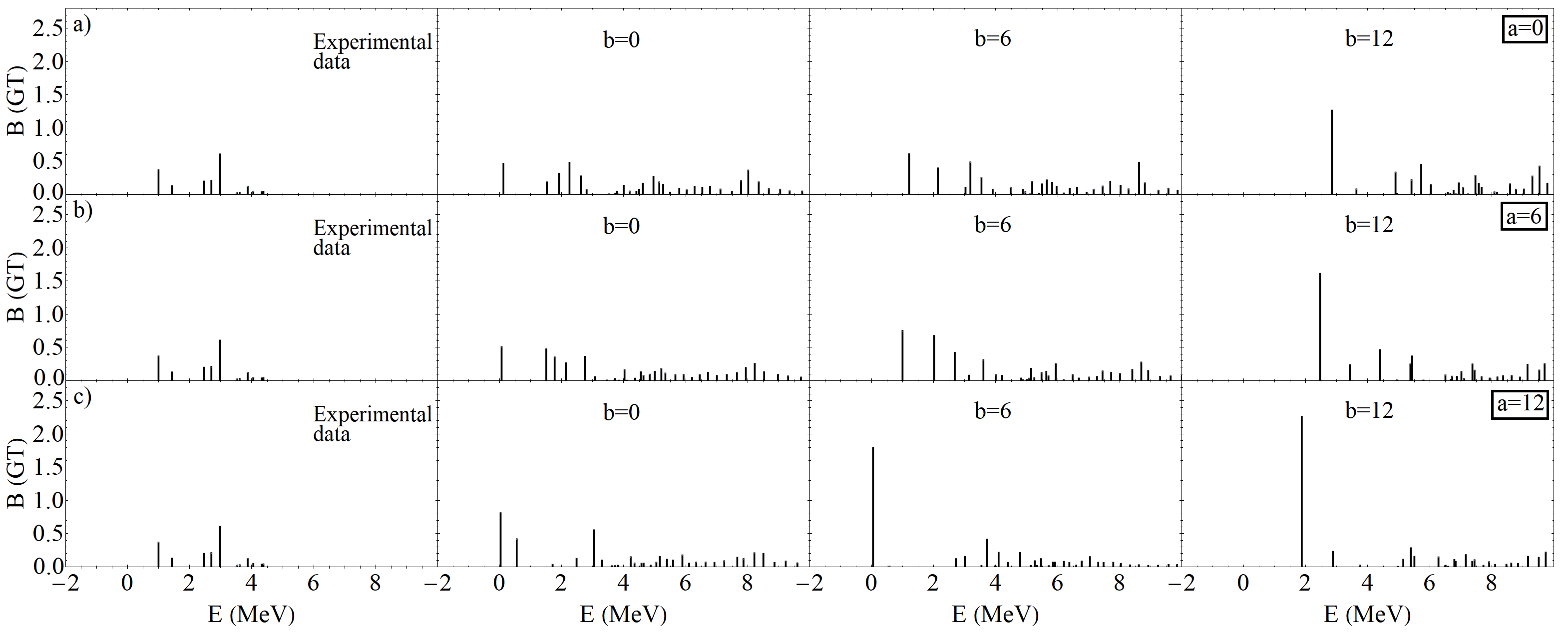}
\caption{Comparison of the experimental \cite{exp46} and theoretical results of Model 2 for  B(GT) transition strengths for $^{46}Ti \rightarrow ^{46}V$ as a function of the isovector pairing strength $b$ and the isoscalar pairing strength $a$, which only acts in the daughter nucleus. }
\end{figure}

The results for Model 1 are especially interesting. For the optimal values of the pairing  strengths $a=b=6$ the lowest $1^+$ state is not the most strongly populated, as is likewise the case experimentally. Earlier we noted that our choice of our optimal parameters was based on getting a good description of the energies of the lowest state, getting a good description of the spread of the rotational band in even--even systems and getting a good description of GT fragmentation. The fragmentation results shown here were critical to our choice of $a=b=6$ with the associated $\chi=-0.065$ parameters. Those chosen in \cite{Stu} are unable to produce the correct fragmentation pattern, as is evident from the $a=b=12$ results shown in the figure.

Other features of the Model 1 results worth noting are that (1) even though we produce an appropriate fragmentation pattern the individual strengths for $a=b=6$ are substantially larger than in the data, (2) the effect of increasing $a$ is, as for other cases, to focus increasing strength in the lowest $1^+$ state while lowering its energy and (3) the effect of increasing $b$ is, as before, to also enhance population of the lowest $1^+$ state but now while lifting its energy.

The results for Model 2 are likewise interesting. Here too there is strong fragmentation of the strength in the optimal $a=b=6$ case, though the lowest state has a bit more strength than the next few. However, the overall strength to these states is in somewhat closer accord with what is seen in experiment. As for Model 1, the effect of increasing $a$ is to focus strength in the lowest $1^+$ state and to lower its energy, while the effect of increasing $b$ is to focus strength on that same state while raising its energy.

\subsubsection{A=48}

Finally, we treat the GT transitions in $ A = 48 $. In this case the relevant decay is $^{48}Ti \rightarrow ^{48}V$, for which both the parent and daughter nuclei have a neutron excess. Thus, in Figs. 27 and 28, where we compare the experimental and calculated transition rates for Models 1 and 2, respectively, the theoretical analysis is only shown as a function of the isovector pairing strength $b$.

\begin{figure}[hbtp]
\centering
\includegraphics[width=1.0\textwidth]{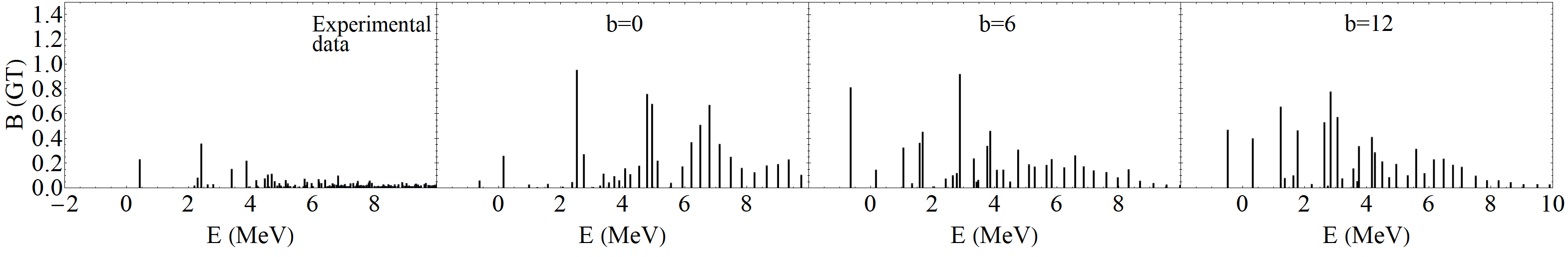}
\caption{Comparison of the experimental \cite{exp48} and theoretical results of Model 1 for  B(GT) transition strengths for $^{48}Ti \rightarrow ^{48}V$ as a function of the isovector pairing strength $b$. Since both nuclei involved have neutron excesses, the isoscalar pairing strength is set to $a=0$ for both.  }
\end{figure}

\begin{figure}[hbtp]
\centering
\includegraphics[width=1.0\textwidth]{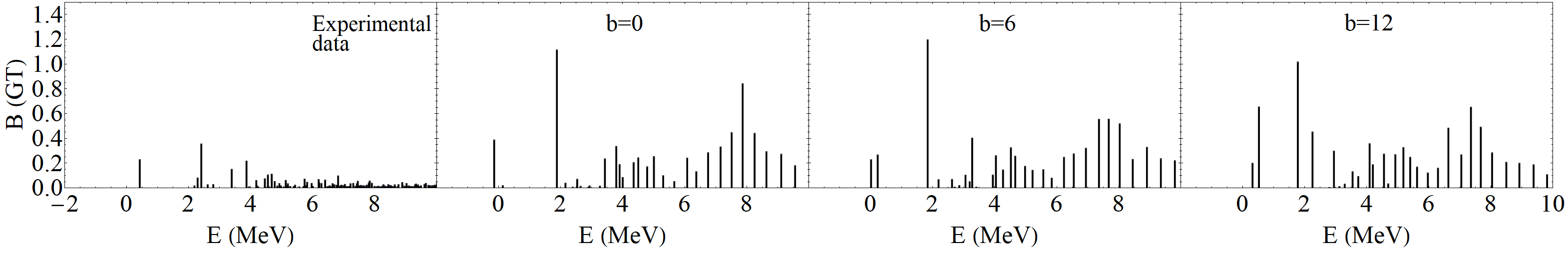}
\caption{Comparison of the experimental \cite{exp48} and theoretical results of Model 2 for  B(GT) transition strengths for $^{48}Ti \rightarrow ^{48}V$ as a function of the isovector pairing strength $b$. Since both nuclei involved have neutron excesses, the isoscalar pairing strength is set to $a=0$ for both.  }
\end{figure}

Note that Model 1 shows the same feature as in the $A=46$ results, namely that for the optimal $a=b=6$ parameters the lowest state is not the most strongly populated, in agreement with experiment. Overall the results show strong fragmentation, although the overall strengths calculated are larger than for experiment. It should also be reiterated that in this case the optimal Hamiltonian does not produce the correct ground state spin and parity.  As $b$ is increased, the overall effect is to increase the level of fragmentation across an increasingly wider range of states.

Model 2 with its optimal parameters is able to obtain the lowest $1^+$ excitation at a reasonable energy, but the level of fragmentation in the data is not as well reproduced as for Model 1. For Model 2 most of the strength appears in a single state near $2~MeV$ in excitation energy, which is where the strongest state lies in the data, but it is several times more strongly populated than any other.  Here too as $b$ is increased, the overall effect is to increase the level of fragmentation across an increasingly wider range of states.

\section{Summary and Conclusions}

In this paper, we explore the effects of proton-neutron pairing on even-mass nuclei in the beginning of $2p1f$ shell. The analysis is done in the framework of the nuclear shell model using two parametrized Hamiltonians that contain not only isoscalar and isovector pairing but also a quadrupole-quadrupole interaction that produces background deformation. The two Hamiltonians differ in the single-particle energies they employ.

We begin the analysis by carrying out a careful search for the optimal values of the parameters of these two Hamiltonians, focusing both on energy spectra {\it and} GT transition properties.  A significant outcome of this part of the analysis is the realization that we need to {\it turn off} isoscalar pairing when dealing with nuclei having a neutron excess to avoid producing the incorrect ground state spin and parity for many such nuclei. Our optimal Hamiltonians are able to achieve good overall fits to experimental data for both spectra and GT decay properties, albeit with the limitations inherent in a relatively simple parametrization. Highlights are an accurate reproduction of the properties of the lowest $1^+$ states, a good description of the overall spread in the energy spectra and a reasonable description of the GT fragmentation pattern.

We then vary the isoscalar and isovector pairing strengths from their optimal values to systematically study how the two pairing modes effect both the energy and GT properties of these nuclei. Our analysis extends from A=42 to A=48. We find that the two pairing modes impact in systematic ways different aspects of the energy spectra and the GT fragmentation pattern. Some of the more interesting observations are that:
\begin {itemize}
\item The isoscalar and isovector pairing modes focus primarily on the odd-J states and the even-J states, respectively. Enhancing the isoscalar strength systematically lowers the first $1^+$ state and expands the set of odd-J states, while leaving the set of even-J states relatively unaffected. In contrast, enhancing the isovector pairing strength expands the even-J part of the spectrum while leaving the odd-J set of states relatively unaffected.
\item Increasing the isoscalar pairing strength in those systems in which it is active ($N =Z$ nuclei)  focuses GT strength on the lowest $1^+$ state while lowering its energy. Increasing the isovector pairing strength, which is always active, also focuses GT strength on the lowest $1^+$ state but raises its energy.
\end{itemize}

\section*{Acknowledgements}
We acknowledge helpful advice of Alfredo Poves on the use of the code ANTOINE and Lei Yang for sharing valuable information.
This work received partial economic support from DGAPA- UNAM project IN109417.


\

\end{document}